\documentclass[journal]{IEEEtran}
\IEEEoverridecommandlockouts
\usepackage{cite}
\usepackage{amsmath,amssymb,amsfonts}
\usepackage{algorithm}
\usepackage{algpseudocode}
\usepackage{graphicx}
\usepackage{textcomp}
\usepackage{xcolor}
\usepackage{fancyhdr}
\usepackage{newtxtext,newtxmath}
\usepackage{subcaption}
\usepackage{siunitx} 
\usepackage{stmaryrd}
\usepackage[T1]{fontenc}
\usepackage{tikz}
\usepackage{bm}
\usepackage{hyperref}
\hypersetup{
    colorlinks,
    linkcolor={black!50!black},
    citecolor={blue!50!black},
    urlcolor={blue!80!black}
}
\hypersetup{hidelinks} 
\usepackage{soul}

\newcommand{\orcid}[1]{\href{https://orcid.org/#1}{#1}} 
\soulregister{\gls}{1}  
\soulregister{\cite}{1} 

\usetikzlibrary{shapes.geometric, arrows}
\tikzstyle{startstop} = [rectangle, rounded corners, minimum width=2cm, minimum height=0.8cm, text centered, draw=black, fill=gray!30]
\tikzstyle{process} = [rectangle, minimum width=2.5cm, minimum height=0.8cm, text width=2.3cm, text centered, draw=black, fill=blue!20]
\tikzstyle{decision} = [diamond, minimum width=1.5cm, minimum height=0.8cm, text width=1.2cm, align=center, text centered, draw=black, fill=green!30]

\tikzstyle{arrow} = [thick,->,>=stealth]

\usepackage{multicol}
\usepackage[acronym]{glossaries}

\makeglossaries
\newacronym{ADAS}{ADAS}{Advanced driver-assistance systems}
\newacronym{FMCW}{FMCW}{frequency modulated continuous wave}
\newacronym{cfar}{CFAR}{constant false alarm rate}
\newacronym{roc}{ROC}{receiver operating characteristic}
\newacronym{SNR}{SNR}{signal-to-noise ratio}
\newacronym{snir}{SNIR}{signal-to-noise-plus-interference ratio}
\newacronym{acc}{ACC}{adaptive cruise control}
\newacronym{RCS}{RCS}{radar cross section}
\newacronym{WGN}{WGN}{white-Gaussian-noise}
\newacronym{DNN}{DNN}{Deep neural network}
\newacronym{FOV}{FOV}{field of view}
\newacronym{SIMO}{SIMO}{single-input-multiple-output}
\newacronym{MIMO}{MIMO}{multiple-input-multiple-output}
\newacronym{LOS}{LOS}{line-of-sight}
\newacronym{NLOS}{NLOS}{non-line-of-sight}
\newacronym{RMSE}{RMSE}{root mean squared error}
\newacronym{RANSAC}{RANSAC}{random sample consensus}
\newacronym{LS}{LS}{least squares}
\newacronym{ML}{ML}{maximum likelihood}
\newacronym{AD}{AD}{autonomous driving}
\newacronym{ISAR}{ISAR}{inverse synthetic aperture radar}
\newacronym{SAR}{SAR}{synthetic aperture radar}
\newacronym{UAV}{UAV}{unmanned aerial vehicle}
\newacronym{UWB}{UWB}{ultra-wide band}
\newacronym{NN}{NN}{neural network}
\newacronym{CNN}{CNN}{convolutional neural network}
\newacronym{FC}{FC}{fully connected}
\newacronym{PRP}{PRP}{perfect reflection point}
\newacronym{MC}{MC}{Monte Carlo}
\newacronym{RA}{RA}{range-azimuth}
\newacronym{LiDAR}{LiDAR}{light detection and ranging}

\def\BibTeX{{\rm B\kern-.05em{\sc i\kern-.025em b}\kern-.08em
    T\kern-.1667em\lower.7ex\hbox{E}\kern-.125emX}}

\newcommand{\Imat}{{\bf{I}}}
\newcommand{\onevec}{{\bf{1}}}
\newcommand{\etavec}{\bm{\eta}}

\begin{document}

\title{A Hybrid Approach for Extending Automotive Radar Operation to NLOS Urban Scenarios}

\author{Aviran~Gal \IEEEmembership{Student Member, IEEE},
        \and
        Igal~Bilik \IEEEmembership{Senior Member, IEEE}%
\thanks{ORCIDs: Aviran Gal \orcid{0009-0003-2110-1388}; Igal Bilik \orcid{0000-0002-0708-4038}.}%
\thanks{Aviran Gal and Igal Bilik are with the School of Electrical and Computer Engineering, Ben-Gurion University of the Negev, Beer Sheva, Israel (e-mails: aviran@post.bgu.ac.il, bilik@bgu.ac.il). This research was partially supported by the Ben-Gurion University of the Negev through the Agricultural, Biological, and Cognitive Robotics Initiative (funded by the Marcus Endowment Fund and the Helmsley Charitable Trust), the Israel Science Foundation under Grant 1895/21, and by the Israeli Smart Transportation Research Center (ISTRC).}%
}

\maketitle

\author{\IEEEauthorblockN{Aviran Gal and Igal Bilik}
\IEEEauthorblockA{\textit{School of Electrical and Computer Engineering}\\
\textit{Ben Gurion University of the Negev}\\
\textit{Beer Sheva, Israel} \\
avirang@post.bgu.ac.il, bilik@bgu.ac.il}
}

\maketitle

\begin{abstract}
Automotive radar is a key component of sensing suites in autonomous driving (AD) and advanced driver-assist systems (ADAS). However, limited line-of-sight (LOS) significantly reduces radar efficiency in dense urban environments. Therefore, automotive radars need to extend their capabilities beyond LOS by localizing occluding and reflective surfaces and non-line-of-sight (NLOS) targets. This work addresses the NLOS target localization challenge by revisiting the NLOS radar signal propagation model and introducing a hybrid localization approach. The proposed approach first detects and localizes reflective surfaces, then identifies the LOS/NLOS propagation conditions, and finally localizes the target without prior scene knowledge, without using Doppler information, and without any auxiliary sensors. The proposed hybrid approach addresses the computational complexity challenge by integrating a physical radar electromagnetic wave propagation model with a deep neural network (DNN) to estimate occluding surface parameters. 
The efficiency of the proposed approach to localize the NLOS targets and to identify the NLOS/LOS propagation conditions is evaluated via simulations in a broad range of realistic automotive scenarios. Extending automotive radar sensing beyond LOS is expected to enhance the safety and reliability of autonomous and ADAS-equipped vehicles.    
\end{abstract}


\section{Introduction}
\gls{ADAS} and \gls{AD} technologies require accurate and reliable information on the vehicle surroundings obtained by the automotive sensing suite~\cite{8828025}. 
Conventionally, automotive sensing suites include cameras, \gls{LiDAR}s, and radars~\cite{ bilik2022comparative}. Automotive radars play a key role in the automotive sensing suite due to their immunity to harsh weather conditions and long operation ranges~\cite{9127857, 9512474,6586127,7485215}. However, in dense urban environments, surrounding obstacles, such as buildings, limit the operational ranges of all conventional sensors and, thus, significantly challenge \gls{ADAS} and \gls{AD} operation~\cite{10149366,8828025,5606708}. 
In addition, the reflective surfaces of the artificial urban obstacles induce multipath propagation phenomena, resulting in ``ghost'' targets that may dramatically degrade automotive radar performance~\cite{10446232, Ref8,6902486,5708185,9455253, 10152506, 9827857}. 

\begin{figure}[!ht]
    \centering
    \includegraphics[width=0.45\textwidth]{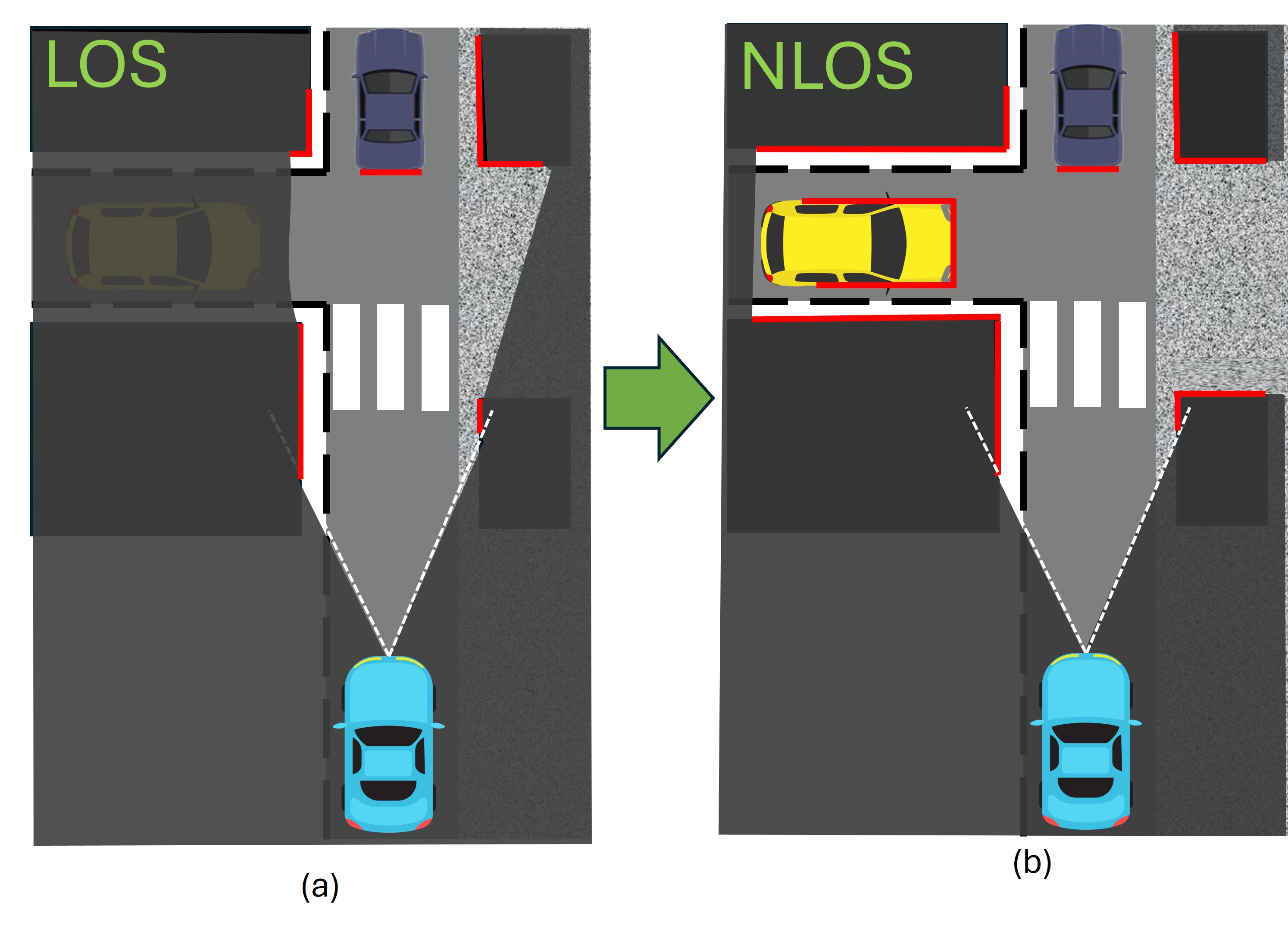}
        \caption[LoF entry]{Illustration of the considered \gls*{NLOS} scenario.
        The yellow vehicle behind the corner is invisible to the radar on the blue vehicle in the conventional \gls*{LOS} scenario in subplot (a).
        The proposed \gls*{NLOS} approach enables detection and localization of the yellow vehicle in subplot (b).}
    \label{fig:nlos_Illustration} 
\end{figure}

Besides these negative aspects, the multipath phenomenon provides an opportunity for radar's \gls*{NLOS} operation. Fig.~\ref{fig:nlos_Illustration} shows the illustrative scenario where the around-the-corner target vehicle \textcolor{black}{(yellow)}, unobservable for any sensor on the host vehicle \textcolor{black}{(blue)} via the conventional \gls*{LOS} propagation conditions \textcolor{black}{in subplot (a)}, can be detected by the radar via the multipath from the reflective surface \textcolor{black}{in subplot (b)}. \textcolor{black}{While \gls*{NLOS} typically induces additional propagation loss, compared to \gls*{LOS} conditions due to multipath reflections, short-range urban scenarios still enable \gls*{NLOS} target detection. In such conditions, shorter propagation distances and larger illuminated reflective surfaces enable \gls*{SNR} at the radar receiver, sufficient for detection and localization of \gls*{NLOS} automotive targets, such as vehicles or motorcycles. \gls*{NLOS} target localization is especially valuable in these short-range urban scenarios, where detecting conventionally invisible threats significantly enhances the host vehicle's situational awareness.}  
Automotive radar operation in such scenarios first requires classification between \gls{LOS} and \gls{NLOS} propagation conditions to identify and mitigate the multipath-induced ``ghost'' targets and accurately localize actual radar targets~\cite{9455253, 9294631, 9764274}. 
This classification requires accurate knowledge of the location and orientation of the surrounding reflective surfaces. Therefore, the problem of estimating the location of reflective surfaces using radar~\cite{10371096, 9636438, 6689216, 9266487, 4290094, 7918866}, \gls{LiDAR}~\cite{1689615,8828025}, and camera~\cite{9114966, 8950766} was recently addressed in the literature. 
Identifying \gls{LOS} and \gls{NLOS} propagation conditions was also studied in the communications framework for channel estimation and power control~\cite{8904260, 6820422, 8968748, 9023805, 8476538, 10233376, 6494741}. 

There is a critical need to extend automotive radar operation capabilities to the \gls{NLOS} propagation conditions. Therefore, various \gls{NLOS} imaging and localization approaches that considered the availability of an accurate {\it a-priori} knowledge of the urban scene geometry have been studied in the literature ~\cite{9074611, 9631228,9978688, 9384308}. 
However, in practice, accurate information on the location of the host vehicle and all surrounding obstacles is unavailable. 
Recently, the \gls{NLOS} targets detection approaches exploiting the targets' motion were proposed in the literature~\cite{8455606, 8640240, 10149715, 8966246, 10371088, 10106370, 8990086, 10296861, 5682038, 10551511}. Some of them used radar's motion to increase the sensor array aperture in the \gls{SAR} framework and thus to achieve high imaging resolution of the vehicle surroundings~\cite{9547412}. Similarly, the \gls{ISAR} framework that exploits the target motion was introduced for \gls{NLOS} imaging of automotive environments~\cite{10522993, 10283172}. A few approaches introduced a target tracking framework to continuously update the target's position~\cite{10644001, 10224179}. 
The \gls{UWB} radar has also been considered for \gls{NLOS} moving targets detection~\cite{9794645, 10285515}.
The bi-static radar framework has been employed to detect \gls{NLOS} target using multiple radar sources in complex environments~\cite{9104299, 10551511}.
Some recent approaches considered sensor fusion and used an auxiliary high-resolution sensor, such as \gls{LiDAR}, to estimate the surrounding obstacles when the  \gls{NLOS} target is located directly behind an obstacle~\cite{9157505, 10152506, 10282816}. 
Recently, the infrastructure-installed intelligent reconfigurable surfaces (IRS) were introduced for \gls{NLOS} target detection~\cite{9468353, 10551511, wei2023multi}. The problem of detecting target transitioning from \gls{NLOS} to \gls{LOS} conditions using tracking approaches was considered in ~\cite{10149684, 10063228}. 
All these approaches consider the availability of auxiliary high-resolution sensors or prior knowledge of the obstacle's location, orientation, and configuration. These considerations are limiting in practice since the scene geometry may not be known, and the availability of the auxiliary sensors is impractical due to the cost constraints of automotive sensing suits for consumer applications.

\gls*{DNN}-based radar processing has been recently introduced in the literature~\cite{danino2024automatic, waldschmidt2021automotive,feintuch2023neural,kang2023deep,DISKIN2024109543, 10124362,feintuch2023neural2}. 
It was proposed for \gls{NLOS} propagation conditions identification~\cite{8968748, 9023805, 8476538, 10233376}, and multipath-induced ghost target mitigation~\cite{9294631, 9764274}. 
\gls{DNN}-based processing was also introduced for reflective surfaces' parameters estimation using measurements from auxiliary  \gls{LiDAR} sensors~\cite{9157505,9294631}. However, all these approaches still conceptually rely on the availability of auxiliary high-resolution sensors and prior knowledge of the obstacle's location, orientation, and configuration.

This work addresses the critical challenge of \gls{NLOS}/\gls{LOS} \gls{MIMO} radar target localization in dense urban environments, operating without {\it a-priori} knowledge of the scene, auxiliary sensors, or Doppler information. Unlike existing methods that rely on additional sensor inputs or prior environment mapping, the proposed approach enables the radar to operate independently in highly obstructed urban environments.
First, we revisit the radar signal model under \gls{NLOS} propagation conditions, incorporating the physical scattering properties of urban structures~\cite{4052607}. Next, we introduce a novel three-stage hybrid approach for LOS/NLOS propagation conditions identification and NLOS target localization. A key innovation of the proposed approach is the combination of the \gls{CNN}-based processing with the physical model of the electromagnetic wave propagation via multipath. 
Thus, the \gls{CNN}-based processing is used to address the computationally complex task of joint estimation of the reflective surface and the \gls{NLOS} target parameters only, while the physical propagation model is used to localize \gls{NLOS} targets via multipath reflections.
The proposed approach consists of three stages: (1) \gls{CNN}-based estimation of reflective surface parameters, (2) identification of the \gls{LOS}/\gls{NLOS} propagation conditions, and (3) accurate target localization using the physical radar signal model. This hybrid framework significantly reduces computational complexity while ensuring robust radar performance in urban environments.

The main contributions of this work are:
\begin{itemize}
    \item A revisited model of \gls{MIMO} radar signal propagation in \gls{NLOS} conditions.
 \item A computationally efficient approach for \gls{NLOS} radar target localization, operating without {\it a-priori} knowledge of the surrounding obstacles, auxiliary sensor, or Doppler information.
    \item A novel hybrid approach that fuses deep learning with the physical model of the radar \gls{NLOS} signal propagation for:
    \begin{itemize}
    \item Accurate \gls{LOS}/\gls{NLOS} propagation conditions identification, 
    \item Precise estimation of reflective surface parameters,
    \item Reliable localization of \gls{LOS}/\gls{NLOS} targets.
    \end{itemize}
\end{itemize}
By extending radar capabilities beyond \gls{LOS} constraints, the proposed approach enhances situational awareness for autonomous driving and \gls{ADAS}, improving safety and operational reliability in complex urban environments.

The following notations will be used throughout this article. 
The super-scripts $\square^{\text{\textit{target}}}$,$\square^{w}$, $\square^{t}$, and $\square^{r}$ denote the target, the reflective surface, and the radar transmitter and receiver, respectively. The sub-scripts $\square_{m_t}$ and $\square_{m_r}$ denote the element indices in the radar transmitter and the receiver arrays, respectively.   
Roman boldface lower-case and upper-case letters denote vectors and matrices, respectively. Nonbold italic letters denote scalars. $\Imat_\nu$ is the $\nu \times \nu$ identity matrix. $\onevec_\nu$ is a vector of ones of length $\nu$. $\left \| \cdot \right \|$ is the $l_2$ norm. $\left( \cdot \right)^T$, $\left( \cdot \right)^H$ represent the transpose and Hermitian transpose operators, respectively. $\operatorname{Re} \left \{ \cdot \right \}$ and $\operatorname{Im} \left \{ \cdot \right \}$ represent the real and imaginary operators. Square brackets, $\left[ \cdot \right]$, denote an element within a vector or matrix.

The remainder of this article is organized as follows. The model of the \gls{MIMO} radar echo received from the reflective surface and the \gls{NLOS} target is revised in Section II. The proposed hybrid approach for the radar \gls{LOS}/\gls{NLOS} target localization is introduced in Section III. The performance of the proposed approach is evaluated in Section IV. Our conclusions are summarized in Section V.

\section{Revisited Model of MIMO Radar NLOS Propagation Conditions} \label{sec:problem_def}
\begin{figure}[!ht]
    \centering
    \includegraphics[width=0.45\textwidth]{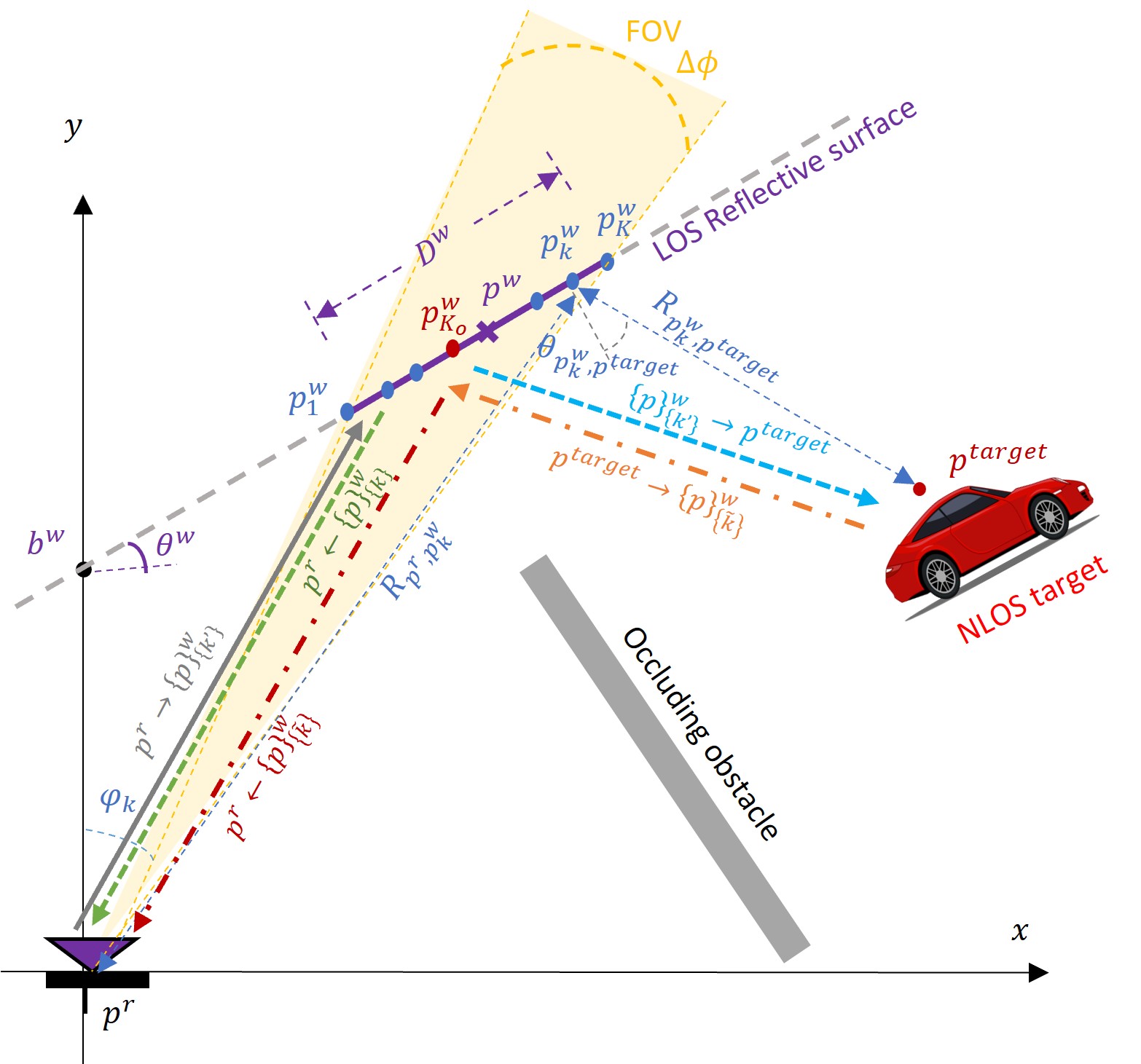}
        \caption[LoF entry]{Schematic representation of the considered \gls*{NLOS} scenario, where the gray obstacle blocks the \gls*{LOS} between the radar in the origin and the red vehicle at $p^{\text{\textit{target}}}$.
        The transmitted radar signal (marked as a solid gray arrow) impinges on the purple reflective surface, modeled as a set of dipole reflectors (marked as blue dots) at the positions, $\left\{p_k^w\right\}_{k=1}^K$. The portion of the impinged energy is reflected back towards the radar (marked as a dashed green arrow). Another portion of the energy (marked as blue dashed arrow) is forward scattered and impinging on the \gls*{NLOS} target (marked in red). The portion of the energy is reflected back towards the reflective surface and impinges on it the second time (marked as orange dash-dot arrow). Finally, the portion for the second time, forward-scattered energy is received at the radar (marked as a red dash-dot arrow).}
    \label{fig:nlos_model} 
\end{figure}

Consider a typical urban scenario in Fig.~\ref{fig:nlos_model}, where the mono-static \gls{MIMO} automotive radar at the origin, $p^r = (x^r,y^r)=(0,0)$, with boresight oriented along the $y$ axis, does not have a line of sight to the stationary \gls{NLOS} \textcolor{black}{point} target at the location, $p^{\text{\textit{target}}}=({x^{\text{\textit{target}}}}, {y^{\text{\textit{target}}}})$ due to the obscuring obstacle. 
In addition, let the reflective surface (marked as a dashed line), with an orientation angle, $\theta^w$, from the $x$-axis, be positioned at a distance, $b^w$, from the radar along its boresight. 
The radar transmits a signal toward the straight reflective surface. The received radar echo is sampled at $n \in {0, 1, \ldots, N-1}$ fast-time instances for the duration of a single transmitted signal, $T_0$.

The reflective surface is modeled as a set of $k=1,\ldots, K$ single-lobe directive radiation antenna elements within a single radar range bin, $\mathbf{p}^w_{K} = \left\{p^w_{k}=(x^w_{k},y^w_{k})\right\}_{k=1}^K$.
The electromagnetic energy reflected from the surface is distributed in space according to some back-scattering pattern, characterized by the reflective surface parameters, such as material, smoothness, and shape~\cite{4052607}. Some of this energy is \textcolor{black}{directly} back-scattered toward the radar, and the $M_r$ antenna elements receive the first radar echo \textcolor{black}{ from the reflective surface}.
Another portion of this energy is forward-scattered toward the \gls{NLOS} target at $p^{\text{\textit{target}}}$. The portion of the energy reflected back from the \gls{NLOS} target impinges the reflective surface \textcolor{black}{a} second time (non-necessarily the same portion of the reflective surface). Some of this energy is forward-scattered toward the radar, and the \textcolor{black}{$M_r$ antenna elements receive the second radar echo from the NLOS target via the two-bounce multipath from the reflective surface.} 

The received radar echo that in the fast-time-receive-channel domain, $\mathbf{X}^r \in \mathbb{C}^{M_r \times N}$, is modeled as a superposition of two reflections: a direct (first) from the reflective surface, $\mathbf{X}^w \in \mathbb{C}^{M_r \times N}$, and the multipath (second), $\mathbf{X}^{\text{\textit{target}}} \in \mathbb{C}^{M_r \times N}$, from the \gls{NLOS} target:
\begin{equation} \label{eq:totalRx}
    \mathbf{X}^r = \mathbf{X}^w + \mathbf{X}^{\text{\textit{target}}} + \mathbf{N}\;,
\end{equation}
where $\mathbf{N}=[\etavec_1 \ldots \etavec_N]$, is the receiver noise matrix with i.i.d. distributed columns, where each, $\left\{\etavec \right\}_{n=1}^N\sim \mathcal{N}^c(\mathbf{0},\,\mathbf{\Gamma})$ is the zero-mean complex additive white Gaussian noise (AWGN) with covariance matrix, $\mathbf{\Gamma}=\sigma_n^{2}\mathbf{I}_{M_r}$, where $\mathbf{I}_{M_r}$ is the identity matrix of the size, $M_r \times M_r$. 
 The radar echoes from the reflective surface, $\mathbf{X}^w$, and the \gls*{NLOS} target, $\mathbf{X}^{\text{\textit{target}}}$, in \eqref{eq:totalRx} are derived in the following subsections.

\subsection{Reflective Surface Geometric Model}
\label{sec:reflective surface}
Consider a single straight reflective surface in Fig.~\ref{fig:nlos_model} of the length, $D^w$, at the distance, $b^w$, from the radar (at the origin), within the radar \gls{FOV}, centered at, $p^{w}=(x^w, y^w)$, and oriented at the angle, $\theta^w$, relatively to the radar boresight, aligned with the $y$-axis. 
 The scatterers' geometric positions along this reflective surface can be modeled as:
\begin{equation} \label{eq:wall_location}
y^w_{k} = x^w_{k} \tan(\theta^w) + b^w\;,\;\;\forall {k}=1,\ldots,{K}\;,
\end{equation}
where the reflector at $p^{w}_{k}$, is located within a single radar range bin of the size, $\Delta r \sim \frac{1}{BW}$, at the range, $R_{p^r, p^{w}_{k}}$, from the radar, at the direction, $\varphi_{k}$, such that:
\begin{equation}
R_{p^r, p^{w}_{k}} \cos(\varphi_{k}) = R_{p^r, p^{w}_{k}} \sin(\varphi_{k}) \tan(\theta^w) +  b^w\;,
 \end{equation}
where \textcolor{black}{$BW$} is the radar bandwidth. 
The reflective surface portion within each range bin is modeled as a single-lobe directive radiation antenna element~\cite{4052607}, at positions, $\mathbf{p}^w_{K} = \begin{bmatrix} p^w_{1} & p^w_{2} & \cdots & p^w_{K} \end{bmatrix}^T$. The beams of these antennas are determined by the reflective surface properties, and their orientation is determined by the illumination directions, $ \{\varphi_{k}\}_{{k}=1}^{K} $, towards the $ \mathbf{p}^w_{K}$ reflectors.
Reflections from different range bins of the reflective surface are considered uncorrelated and are processed independently.

\subsection{The Radar Echo From the Reflective Surface} 
Consider the signal from the $m_t$th, $\forall m_t=1,..., M_t$ transmitter of the radar transmit array, located at, $p^r_{m_t}$, illuminating the reflective surface, represented as a set of reflection points, $p^w_{k}\;,\forall {k}=1,\ldots, K$, modeled as dipole antenna with a corresponding beam pattern~\cite{4052607}, where a single dipole antenna is assigned to each range-DOA cell. 

Assuming the narrow-band radar and far-field propagation conditions from the radar to the reflective points, $\left\{p^w_{k}\right\}_{k=1}^{K}$, the radar echo from the entire reflective surface at the time instance $n$, is modeled as a superposition of $K$ radar echoes as:
\begin{equation} \label{eq:wall_model}
    [\mathbf{X}^w]_n = \sum_{{k}=1}^{K} \Tilde{\alpha}_{k}^w L^2_{p^{r},p^{w}_{{k}}}  {\mathbf{a}}_r(\varphi_{k}) {\mathbf{a}}_t^T(\varphi_{k}) \mathbf{s}( \frac{T_0}{N} n - \tau_{k}) ,
\end{equation}
where $\Tilde{\alpha}_{k}^w = \sigma^w_{k} S^b_{{k}}$, $\sigma^w_{k}$ is Rayleigh-distributed \gls{RCS}, where  $ S^b_{k} = (1 - \Lambda) $ is the backscattering reflection function~\cite{4052607} towards the radar from the ${k}$th scatterer $ p^w_{k}$, where $ \Lambda \in [0,1] $ is the backscattering reflection ratio, $L^2_{p^r,p^w_{{k}}}$ is the propagation loss in the path from the radar to the ${k}$th reflector, $p^w_{{k}}$, 
$\mathbf{a}_r(\varphi_{k}) \in \mathbb{C}^{M_r}$ and $\mathbf{a}_t(\varphi_{k}) \in \mathbb{C}^{M_t}$ are the receive and the transmit steering vectors, respectively, towards direction $\varphi_{k}$~\cite{5236885}, $\mathbf{s}_t(t)$ is the transmit signal vector, $\tau_{{k}} = \frac{2}{c} R_{p^r, p^w_{{k}}, p^r}$ is a two-way propagation time delay from the radar to the $p^w_{{k}}$ reflector at the round-trip range, $R_{p^r, p^w_{{k}}, p^r}$.
\textcolor{black}{Notice that the model in~\eqref{eq:wall_model} holds for any radar transmit waveform.}  

\subsection{The Radar Echo from the NLOS Target} \label{sec:target}
Let the forward-scattering reflections from the reflective points, $\left\{p^w_{k'}\right\}_{k'=1}^{K'}$ on the reflective surface illuminate the \gls{NLOS} target, with significantly smaller dimensions than the reflective surface length, at the location, $p^{\text{\textit{target}}}$. The radar echo reflected back from the anisotropic target toward the same reflective surface illuminates possibly a different subset of reflective points, $\left\{p^w_{\Tilde{k}}\right\}_{\Tilde{k}=1}^{\Tilde{K}}$. The forward-scattered reflections from these points are received by the sensor array at the radar receiver. The echo at the $m_r$th radar receiver array element is modeled as a superposition of signals propagating via the paths, $p^r_{m_t} \shortrightarrow {p}^w_{k'} \shortrightarrow p^{\text{\textit{target}}} \shortrightarrow {p}^w_{\Tilde{k}} \shortrightarrow p^r_{m_r}, \forall k='1,\ldots, K', \tilde{k}=1,\ldots,\tilde{K}$. 
\textcolor{black}{In this work, only first- and second-order reflections are considered, and higher-order multipaths from other surrounding structures are assumed to be negligible, consistent with practical scenarios where such contributions appear below the radar detection threshold.}

The radar echo at the receiver sensor array at the time instance $n$, $[\mathbf{X}^{\text{\textit{target}}}]_{n} \in \mathbb{C}^{M_r}$, is:
\begin{align} \label{eq:mimo_target}
    [\mathbf{X}^{\text{\textit{target}}}]_{n} &= \sigma^{\text{\textit{target}}}
    \sum_{{\Tilde{k}}=1}^{\Tilde{K}} \alpha_{\Tilde{k}} \mathbf{a}_r(\varphi_{\Tilde{k}}) \sum_{k'=1}^{K'} \alpha_{k'} \nonumber \\
     &\cdot \mathbf{a}_t^T(\varphi_{{k'}}) \mathbf{s}\left(\frac{T_0}{N} n - \tau_{r,k'} - \tau_{k',t} - \tau_{t, \Tilde{k}} -\tau_{\Tilde{k}, r} \right) \;,
\end{align}
where $\mathbf{a}_r(\varphi_{\Tilde{k}}) \in \mathbb{C}^{M_r}$  and $\mathbf{a}_t(\varphi_{k'}) \in \mathbb{C}^{M_t}$ are the receive and transmit steering vectors towards direction $\varphi_{\Tilde{k}}$ and $\varphi_{k'}$, respectively~\cite{5236885}, $\tau_{r,{k'}} = \frac{2}{c} R_{p^r, p^w_{k'}}$, $\tau_{{k'}, t} = \frac{2}{c} R_{p^w_{{k'}},p^{\text{\textit{target}}}}$, $\tau_{t,\Tilde{k}} = \frac{2}{c} R_{p^{\text{\textit{target}}}, p^w_{\Tilde{k}}}$, $\tau_{\Tilde{k}, r} = \frac{2}{c} R_{p^w_{\Tilde{k}}, p^r}$, are the components of the one way propagation time delay in path, $p^r \shortrightarrow {p}^w_{k'} \shortrightarrow p^{\text{\textit{target}}} \shortrightarrow {p}^w_{\Tilde{k}} \shortrightarrow p^r$. 
The propagation coefficients are, $\alpha_{k'} = \sigma^w_{k'} \gamma_{k'} L_{p^r,p^w_{k'},p^{\text{\textit{target}}}}$ and $\alpha_{\Tilde{k}} = \sigma^w_{\Tilde{k}} \gamma_{\Tilde{k}} L_{p^{\text{\textit{target}}},p^w_{\Tilde{k}},p^r}$ where $\sigma^w_{\Tilde{k}}$ and $\sigma^w_{k'}$ are the reflective surfaces' \gls{RCS} and $\sigma^{\text{\textit{target}}}$ is the target \gls{RCS} Rayleigh complex scattering coefficients, where $\gamma_{\Tilde{k}}$ and $\gamma_{k'}$ are the functions of the incidence and scattering directions, according to~\cite{4052607}:
\begin{eqnarray}
\label{eq:scattering}
    \gamma^2_{k'} &= \Lambda^2\left(\frac{1+\cos\left(\varphi_{k'} + \theta^w -\varphi_{p_{k'}^w,p^{\text{\textit{target}}}}\right)}{2}\right)^{\psi^w}\;,\\
    \gamma^2_{\Tilde{k}} &= \Lambda^2\left(\frac{1+\cos\left(\varphi_{\Tilde{k}} + \theta^w -\varphi_{p_{\Tilde{k}}^w,p^{\text{\textit{target}}}}\right)}{2}\right)^{\psi^w}\;,
\end{eqnarray}
where $\psi^w$ determines the width of the reflected beam. The path losses for the paths $p^r \shortrightarrow {{p}}^w_{k'} \shortrightarrow p^{\text{\textit{target}}}$ and $p^{\text{\textit{target}}} \shortrightarrow {{p}}^w_{\Tilde{k}} \shortrightarrow p^r$, are $L_{p^r,p^w_{k'},p^{\text{\textit{target}}}}$ and $L_{p^{\text{\textit{target}}},p^w_{\Tilde{k}},p^r}$, respectively. \textcolor{black}{Notice that the model in~\eqref{eq:mimo_target} holds for any radar transmit waveform.}  

\subsection{Target Location} \label{sec:target_est}
The target location can be modeled using the considered scenario geometry from Subsections A - C as:
\begin{align} \label{eq:NLOS_target}
   x^{\text{\textit{target}}} &= R_{p^r, p^w_{K_o}} \sin ({\varphi}_{K_o}) + {R}_{p^w_{K_o}, p^{\text{\textit{target}}}} \sin (2 \theta^w +  {\varphi}_{K_o}) , \nonumber \\
    y^{\text{\textit{target}}} &= R_{p^r, p^w_{K_o}} \cos (\varphi_{K_o}) - {R}_{p^w_{K_o}, p^{\text{\textit{target}}}} \cos (2 \theta^w + \varphi_{K_o}) ,
\end{align}
where $p^w_{K_o}$ is the \gls{PRP} on the reflective surface where the incident wavefront obeys the Snell’s law, such that the incidence angle, $\varphi_{K_o}$, equals the reflection angle, $\theta_{p^w_{K_o},p^t}$, relative to the local surface normal. At the \gls{PRP}, $\gamma^2_{K_o} = \Lambda^2$, 
$R_{p^r, p^w_{K_o}}$ is the range to this \gls{PRP} from the radar, along the path $p^r \shortrightarrow {p}^w_{{K_o}}$, and ${R}_{p^w_{K_o}, p^{\text{\textit{target}}}}$ is the range from the \gls{PRP} to the target, along the path ${p}^w_{{K_o}} \shortrightarrow p^{\text{\textit{target}}}$. The target location in Cartesian coordinates, $(x^{\text{\textit{target}}}, y^{\text{\textit{target}}})$,  relative to the radar, is proposed to be estimated using radar echoes from both the reflective surface, and the target, received via the direct and reflective paths, respectively.

\begin{figure*}
    \centering 
    \includegraphics[width=\textwidth]{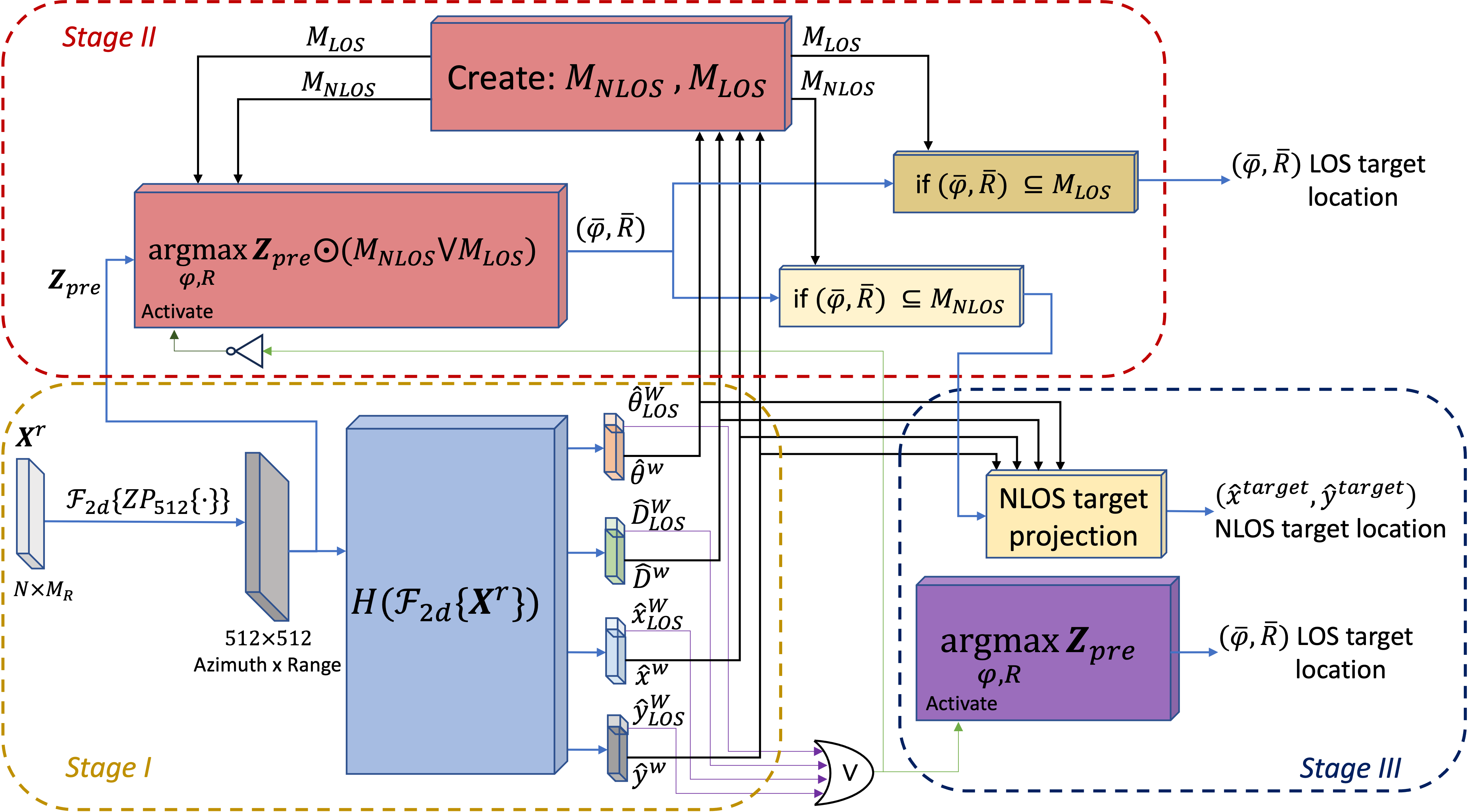}
    \caption[LoF entry]{Schematic representation of the proposed hybrid approach with three-stage processing for the \gls{NLOS}/\gls{LOS} target localization.
    In Stage I, the reflective surface within the scene is detected, and its parameters are estimated, as detailed in Section~\ref{sec:wall_e}.
    In Stage II,  the classification between NLOS and LOS propagation conditions is performed, as described in Section~\ref{sec:NLOS_LOS_det}.
    The \gls{NLOS}/\gls{LOS} target is localized in Stage III, as detailed in Section~\ref{sec:target_e}.
    The output of the proposed approach is the estimated target location, $(\hat{x}^{\text{\textit{target}}}, \hat{y}^{\text{\textit{target}}})$.}
    \label{fig:algo}
\end{figure*}

\section{The Proposed Approach}
\begin{figure}[ht]
\centering
\resizebox{\columnwidth}{!}{
\begin{tikzpicture}[node distance=1.5cm]

\node (start) [startstop] {$\mathbf{X}^r$};
\node (step1) [process, below of=start, yshift=0.2cm] {Stage I};
\node (step1_dec1) [decision, below of=step1, yshift=-0.8cm, align=center] {Reflective Surface\\Exist?};
\node (step3_los) [process, below of=step1_dec1, yshift=-1.2cm] {Stage III (LOS Target)};
\node (step2) [process, right of=step1_dec1, xshift=2.5cm] {Stage II};
\node (step2_dec2) [decision, below of=step2, yshift=-1.2cm] {Target Propagation?};
\node (step3_nlos) [process, right of=step2_dec2, xshift=2.5cm] {Stage III (NLOS Target)};

\node (los_target) [startstop, below of=step3_los, yshift=0.2cm] {LOS Target Location};
\node (nlos_target) [startstop, below of=step3_nlos, yshift=0.2cm] {NLOS Target Location};

\draw [arrow] (start) -- (step1);
\draw [arrow] (step1) -- (step1_dec1);
\draw [arrow] (step1_dec1) -- node[yshift=-6pt, xshift=10pt, anchor=south] {No} (step3_los);
\draw [arrow] (step1_dec1) -- node[yshift=6pt, xshift=5pt, anchor=east] {Yes} (step2);
\draw [arrow] (step2) -- (step2_dec2);
\draw [arrow] (step2_dec2) -- node[yshift=6pt,xshift=-8pt, anchor=west] {LOS} (step3_los);
\draw [arrow] (step2_dec2) -- node[yshift=6pt,xshift=17pt, anchor=east] {NLOS} (step3_nlos);
\draw [arrow] (step3_los) -- (los_target);
\draw [arrow] (step3_nlos) -- (nlos_target);

\end{tikzpicture}
}
\caption{Decision tree of the proposed approach.}
\label{fig:decision_tree}
\end{figure}
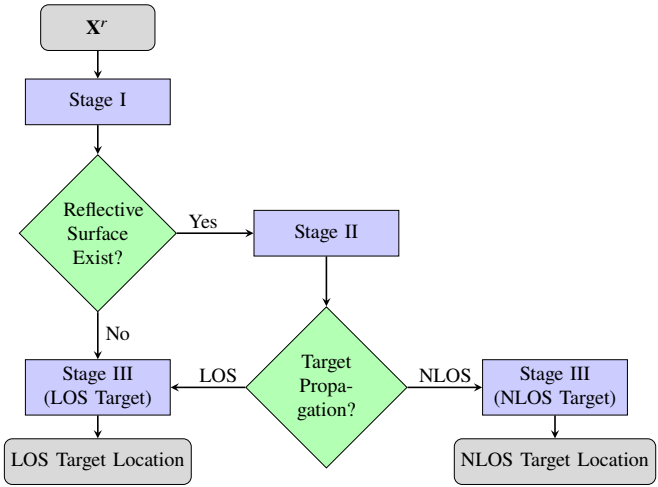

This section summarizes the proposed hybrid approach for the \gls{NLOS}/\gls{LOS} target detection, identification, and localization, schematically shown in Fig.~\ref{fig:algo}. 
The decision tree in Fig.~\ref{fig:decision_tree} shows the logic of the proposed approach.
One of the major challenges of \gls{NLOS} target detection, identification, and localization is the high dimensionality of the problem, which is associated with the need to estimate both the target and the reflective surface parameters jointly. Conventional parameter estimation approaches are infeasible for the considered problem. Therefore, this work proposes a hybrid approach that combines the physical model of the radar signal propagation with the DNN-based processing that efficiently addresses the problem's high computational complexity.
The proposed hybrid approach consists of three stages. Following the logic in Fig.~\ref{fig:decision_tree}, in the first stage, summarized in Subsection~\ref{sec:wall_e}, the reflective surface within the radar \gls{FOV} is detected, and its parameters are estimated. 
In the second stage, summarized in Subsection~\ref{sec:NLOS_LOS_det}, the \gls{NLOS} or the \gls{LOS} propagation conditions are identified. Finally, the target location is estimated in the third stage, detailed in Subsection~\ref{sec:target_e}.

\subsection{Stage I: Reflective Surface Detection and Parameters Estimation} \label{sec:wall_e}
The proposed approach for the reflective surface detection and estimation of its parameters is detailed in Algorithm~\ref{algo:stage_1}.
Let the unknown parameters determining a finite reflective surface be the reflective surface center, length, and orientation angle:
\begin{equation}
    \hat{\mathbf{\xi}}^w = [\hat{x}^w, \hat{y}^w, \hat{D}^w, \hat{\theta}^w]\;. \label{reflective surfaceParam}   
\end{equation}
The main idea of the proposed approach is to leverage the convolutional filters in a CNN to estimate the object's center-of-mass and to detect its presence within the radar \gls{FOV}.
The high complexity of the mapping between the \gls{NLOS} target location and the received radar range-DOA map motivates the proposed \gls{DNN}-based approach leveraging the {\it EfficientNet $b_1$} estimator~\cite{tan2020efficientnet}. This architecture is particularly well-suited for processing the radar range-DOA representation, where, similarly to natural images, the required information can be extracted from the spatial structure.

According to the radar signal propagation model, introduced in Section~\ref{sec:problem_def}, the radar echo from the reflective surface can be \textcolor{black}{represented as a superposition of echoes from a large unknown number of reflectors. Estimating the parameters of these reflectors is a high-dimensional and highly nonlinear problem, which is infeasible for conventional estimation approaches.} Therefore, the proposed approach combines this model with the \gls{DNN}-based processing to estimate the reflective surface parameters from the received radar range-angle measurements. 

The proposed approach, schematically shown in Fig.~\ref{fig:wall_param_NN}, performs the nonlinear mapping, $\mathbf{X}^r \shortrightarrow \hat{\mathbf{\xi}}^w$, from the radar receiver measurements in~\eqref{eq:wall_model} into the vector of parameters in~\eqref{reflective surfaceParam} as:
\begin{equation}
    \hat{\mathbf{\xi}}^w = \mathbf{H}(\mathbf{X}^r) \;.
\end{equation}
The following subsections detail the radar echoes pre-processing, the considered \gls{DNN} architecture, and the algorithm for reflective surface parameters estimation.

\subsubsection{Pre-processing}
The proposed \gls{DNN} architecture requires pre-processing of the received radar echoes, $\mathbf{X}^r$, in order to create a \gls{RA} high-resolution map at its input:
\begin{equation} \label{eq:z_pre}
    \mathbf{Z} = \mathcal{F}_{2d} \left(P_{512}\left(\mathbf{X}^r\right)\right) \in \mathbb{C}^{512 \times 512}\;,
\end{equation}
where $\mathcal{F}_{2d}(\cdot)$ is the 2D-FFT operator and $P_{512}(\cdot)$ denotes zero padding to $512$ in each \gls{RA} dimension. The output of this stage is a complex matrix, which is vectorized into $\mathbf{Z}_{pre} = [\operatorname{Re}\{\mathbf{Z}\}, \operatorname{Im}\{\mathbf{Z}\}, |\mathbf{Z}|] \in \mathbb{R}^{512 \times 512 \times 3}$, down-sampled and bilinear interpolated, $\mathbf{Z}_{in} \in \mathbb{R}^{240 \times 240 \times 3}$ to match the input dimensions of the \gls{CNN} architecture.
\textcolor{black}{Notice that although the real and imaginary parts of the \gls{RA} map contain all the information, the magnitude $|\mathbf{Z}|$ was used as a third channel for input to the \gls{CNN}. Although theoretically redundant, the empirical experiments demonstrated that this additional channel improved performance, likely due to its interpretable structure and complementary distribution. Following this observation, the phase information, $\angle\mathbf {Z}$, was considered instead of the real or imaginary parts. However, this additional phase channel degraded the performance, likely due to normalization mismatches and training instability.}

\subsubsection{The CNN Architecture}
This work \textcolor{black}{exploits the observation} that the relative geometry of the \gls{NLOS} target with the reflective surface induces a unique range-DOA map at the radar receiver that can be used for the \gls{NLOS} target localization.
The \gls*{DNN} estimator performing the following mapping:
\begin{equation} \label{eq:eff_net}
    \mathbf{\zeta} = \mathbf{C}(\mathbf{Z}_{in}) \;,
\end{equation}
where $\mathbf{C}(\cdot)$ denotes the {\it EfficientNet $b_1$} network, and the output features vector, $\mathbf{\zeta}$,  contains the information on $\hat{\mathbf{\xi}}^w$. In the considered here \gls{DNN} architecture, the last classifier layer is removed from the original {\it EfficientNet $b_1$}, to interconnect the \gls{CNN} with each $\mathcal{P}^{\theta/D/x/y}$ block. This modification leads to $6,513,184$ learnable parameters of {\it EfficientNet $b_1$} network.

 \begin{figure*}[htp]
    \centering 
    \includegraphics[width=0.7\textwidth]{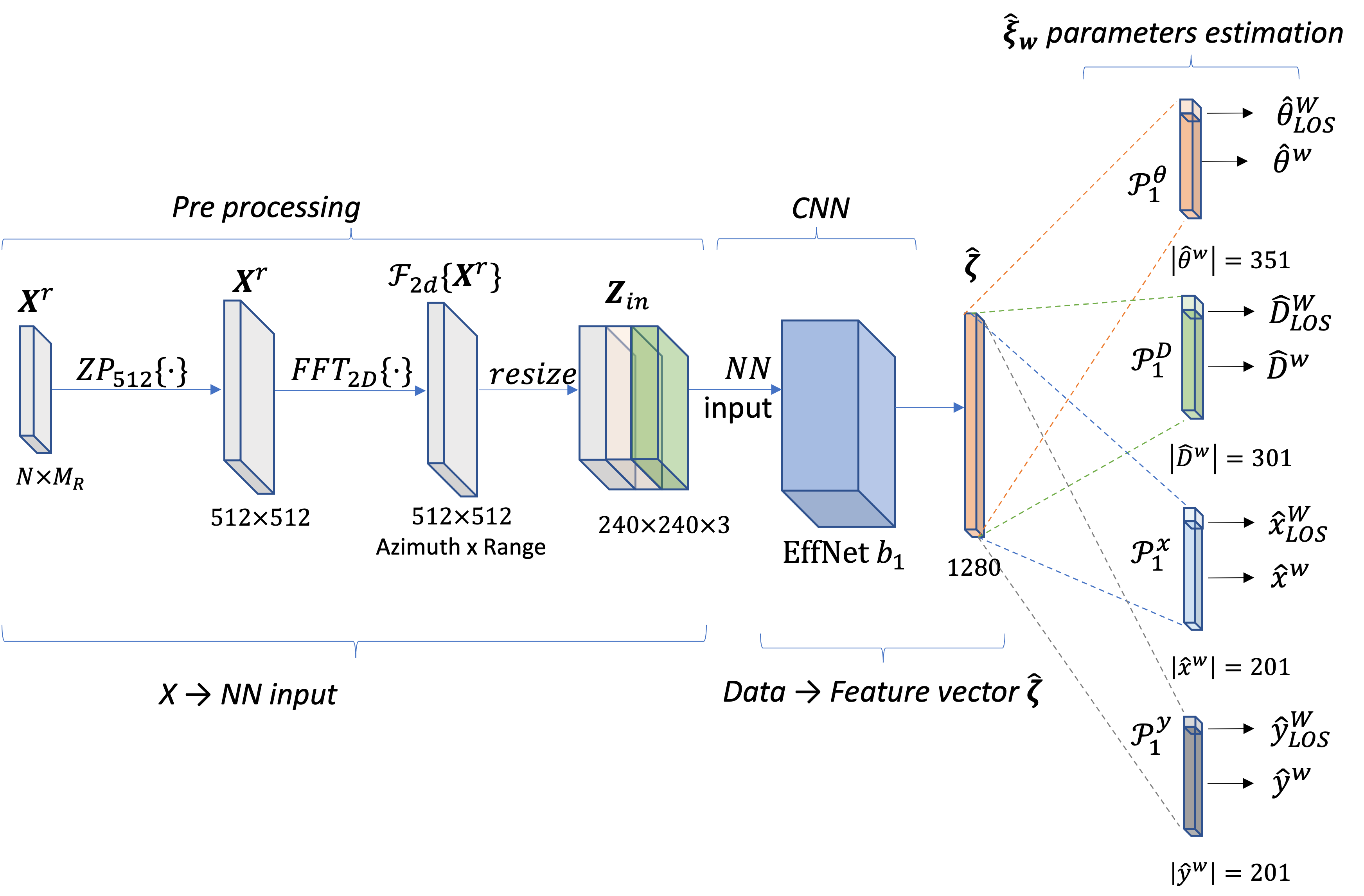}
     \caption{Proposed \gls{NN} architecture for reflective surface parameter estimation in Stage I (Algorithm 1). It includes pre-processing (zero padding, 2D FFT, resizing), a \gls{CNN} ({\it EfficientNet $b_1$}) for feature extraction, and $4$ fully connected layers for parameter estimation.}
         \label{fig:wall_param_NN}
\end{figure*}

\begin{table}[!ht]
    \centering
       \caption{Parameters of the \gls{NN} architecture in Fig. 4, used in Stage I for reflective surface parameters estimation.}
       \label{table:FC_layers_param}
     \begin{tabular}{l l l S[table-format=7.0, group-four-digits=true, group-separator={,}]}
    \hline\hline
        Layer & {$D_1 \times D_2$} & $\sigma$ function & {\#Parameters}  \\[1.0ex]  \hline\hline 
        \rule{0pt}{2.5ex}$\mathbf{C}(\cdot)$ & -- & {\it EfficientNet $b_1$} & 6513184  \\ [1.0ex] \hline
        \rule{0pt}{2.5ex}$\mathcal{P}^{\theta}$ & $1280 \times 351$ & \textit{LeakyReLU} & 449631  \\ [1.0ex] \hline
        \rule{0pt}{2.5ex}$\mathcal{P}^{D}$ & $1280 \times 301$ & \textit{LeakyReLU} & 385581  \\ [1.0ex]\hline
        \rule{0pt}{2.5ex}$\mathcal{P}^{x}$ & $1280 \times 201$ & \textit{LeakyReLU} & 257481  \\ [1.0ex]\hline
        \rule{0pt}{2.5ex}$\mathcal{P}^{y}$ & $1280 \times 201$ & \textit{LeakyReLU} & 257481  \\ [1.0ex]\hline\hline
        \rule{0pt}{2.5ex}Total& -- & -- & 7863358 \\ [1.0ex]\hline
    \end{tabular}
    \end{table}
    
\subsubsection{Reflective Surface Detection and Parameters Estimation}
The problem of reflective surface detection and reconstruction involves estimating the parameters, $\hat{\mathbf{\xi}}^w$. According to the proposed approach, the reflective surface parameters are estimated in the third block in Fig.~\ref{fig:wall_param_NN}. 
Dimensions of each parameter in $\hat{\mathbf{\xi}}^w$ are reduced by $4$ parallel \gls{FC} layers, as defined in stage $5$ in Algorithm~\ref{algo:stage_1}, with parameters, specified in Table~\ref{table:FC_layers_param}. In Algorithm~\ref{algo:stage_1}, the weight matrix, $\mathbf{W} \in \mathbb{R}^{D_1 \times D_2}$ and the bias vector, $\mathbf{b} \in \mathbb{R}^{D_2}$ are optimized during the training process, and $\sigma(\cdot)$ is an element-wise nonlinear activation function. The size of each parameter in the vector, $\hat{\mathbf{\xi}}^w$, defines the achievable estimation resolution. Each FC layer has two objectives, which are met simultaneously, as described in stage $6$ of Algorithm~\ref{algo:stage_1}: 
\begin{enumerate}
    \item Identify the reflective surface in the first element.
    \item Estimate the surface parameters with the rest of the elements.
\end{enumerate}

During training, the {\it cross entropy} loss function is used to perform the derivative, instead of the $\operatorname*{arg\,max}_{(\cdot)} \mathcal{P}^{\theta/D/x/y}(\cdot)$. The total number of learnable parameters in this NN architecture is $7,863,358$.

\subsection{Stage II: LOS/NLOS Target Propagation Conditions Identification} \label{sec:NLOS_LOS_det}
The reflective surface, detected and localized in Stage I, does not necessarily obscure the radar target.
This subsection introduces the \gls{NLOS} propagation conditions identification approach summarized in Algorithm~\ref{algo:stage_2} by segmenting the radar \gls{FOV} into \gls{LOS} and \gls{NLOS} regions, according to $\hat{\xi}^w$.

The main idea of this stage is to formulate this problem as a classification between: 
\begin{align}
    I_0 &: \textrm{\gls{LOS} propagation conditions}\;, \nonumber\\
    I_1 &: \textrm{\gls{NLOS} propagation conditions}\;.
\end{align} 

\begin{figure}[htp]
    \centering 
    \includegraphics[width=0.48\textwidth]{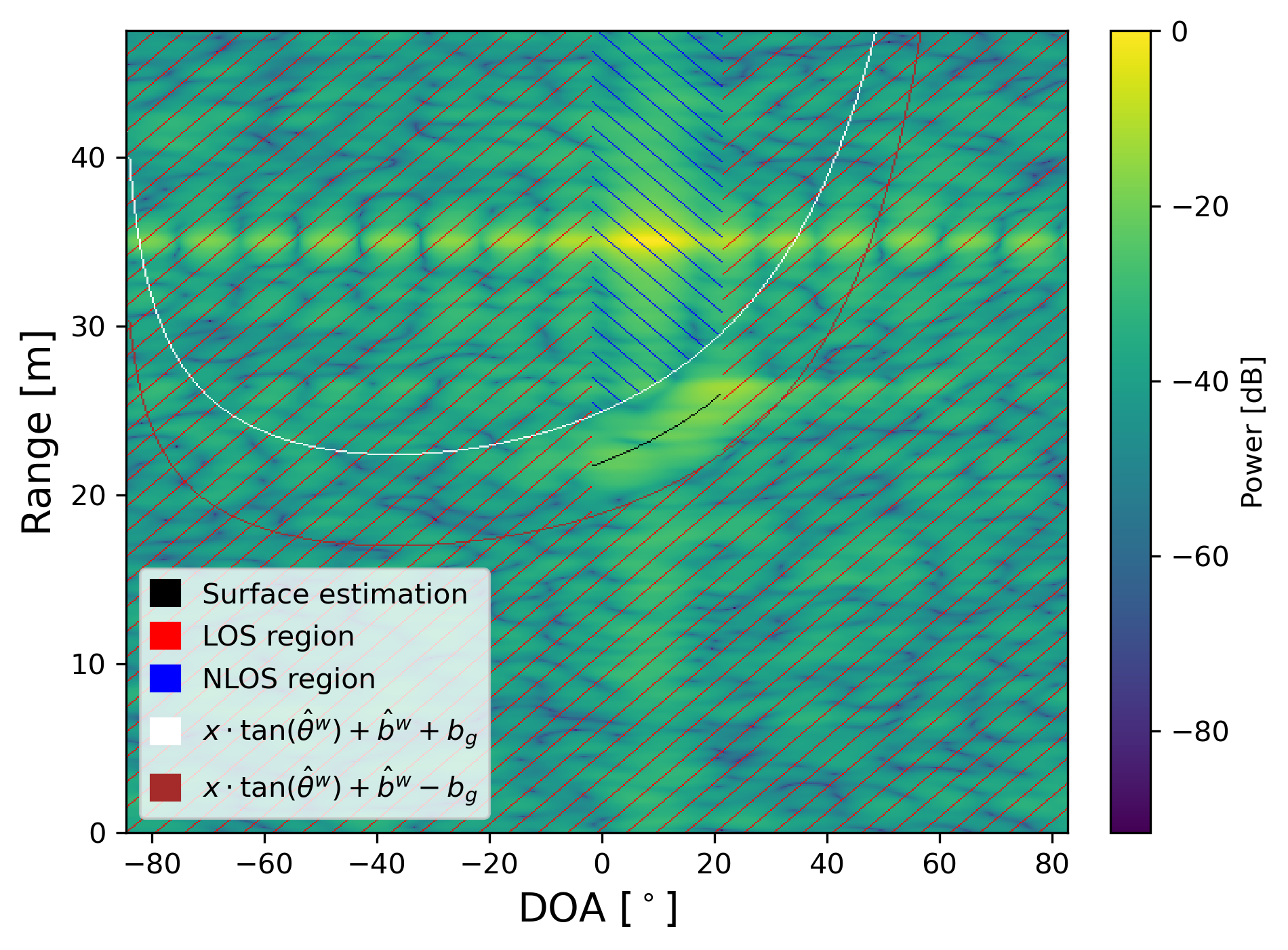}
     \caption[LoF entry]{Exemplary normalized \gls*{RA} map, $\mathbf{Z}$, obtained as a 2D-FFT on the received radar echo, $\mathbf{X}^r$, which includes reflections from the surface, \gls{NLOS} target, and the additive noise.
     
     \ \ \ In this scenario, the following parameters were considered for the reflective surface: ${x}^w = 3.3 m, {y}^w = 23.3 m, {D}^w = 8.8m, \theta^w=25^\circ$, $\text{SNR}^w = 30\,\mathrm{dB}$, and for the  NLOS target: $\varphi_{K_o}=6.3^\circ, R_{p^r, p^w_{K_o}} = 22.9 m, {R}_{p^w_{K_o}, p^{\text{\textit{target}}}} = 11.9m$, $\text{SNR}^{\text{\textit{target}}} = 50\,\mathrm{dB}$.
     
     \ \ \ The LOS region is marked as red, and the estimated reflective surface, marked in black, creates the NLOS region, marked as blue. The peak value is set to $0$ dB, and the colorbar represents the power level in dB, relative to this peak.}
         \label{fig:RA_wall_target_with_lines}         
\end{figure}

If a reflective surface is detected within the radar FOV in Stage I, the radar \gls{FOV} is segmented into \gls{LOS} and \gls{NLOS} regions according to the binary masks, $\mathbf{M}_{NLOS}$ and $\mathbf{M}_{LOS}$. The finite-length flat reflective surface within the radar \gls{FOV}, modeled in \eqref{eq:wall_location}, can be estimated as:
\begin{equation} \label{eq:surface_line_algo}
    y = x \tan(\hat{\theta}^w) + \hat{b}^w \pm b_g\;,
\end{equation}
where $\hat{b}^w = \hat{y}^w - \hat{x}^w \tan(\hat{\theta}^w)$. \textcolor{black}{For each \gls*{RA} cell within the radar’s \gls*{FOV}, a \gls*{LOS} between the radar and this cell is evaluated by assessing the direct geometric path from the radar to the cell. If this path intersects with the estimated reflective surface in~\eqref{eq:surface_line_algo}, the cell is classified as \gls*{NLOS}. Otherwise, it is considered \gls*{LOS}. This classification process results in two binary masks, $\mathbf{M}_{\text{LOS}}$ and $\mathbf{M}_{\text{NLOS}}$, corresponding to the visible and occluded regions, respectively.}
 The estimated reflected surface is then represented in the polar coordinates, aligned with the $\mathbf{Z}$ matrix. 
 \textcolor{black}{Fig.~\ref{fig:RA_wall_target_with_lines} shows the normalized \gls*{RA} map, where the peak value is set to $0$ dB, and the color represents the power values in dB, relative to it.} It is assumed that the target can only be at a distance larger than $b_g$[m] from the reflective surface. The estimated reflective surface is marked with a black line, expressions in~\eqref{eq:surface_line_algo} are marked as white and brown, and the LOS and NLOS regions are shaded with red and blue line marks, respectively. Notice that, considering the challenging static scenario, where the target Doppler information can not be used to discriminate target echo from the static reflective surface, the sidelobes of the reflective surface echo can mask the target echo. 
The union of $\mathbf{M}_{LOS}$ and $\mathbf{M}_{NLOS}$ masks the reflective surface and its sidelobes in the range-DOA domain, $\mathbf{Z}$. Using this masking, the target location within the entire \gls{NLOS}/\gls{LOS} region can be estimated using the remaining radar range-DOA map as:
{\textcolor{black}{
\begin{equation} \label{eq:argmax_Z}
    \left\{\bar{\varphi}^*, \bar{R}^*\right\} = \operatorname*{arg\,max}_{\bar{\varphi},  \bar{R}} |\mathbf{Z} \odot (\mathbf{M}_{LOS} \lor \mathbf{M}_{NLOS})|\;.
\end{equation}
where $\lor$ denotes the element-wise logical OR operation between the LOS and NLOS masks.}
These estimates can be used to decide on $I_0$ or $I_1$ hypothesis as:
\begin{align}
\left\{\bar{\varphi}^*, \bar{R}^*\right\}&\in \mathbf{M}_{LOS}\rightarrow I_0\;,\nonumber\\
\left\{\bar{\varphi}^*, \bar{R}^*\right\}&\in \mathbf{M}_{NLOS}\rightarrow I_1\;.\nonumber
\end{align}
Notice that the LOS target in the presence of the multipath, received via $p^r \shortrightarrow p^{\text{\textit{target}}} \shortrightarrow {p}^w_{\Tilde{k}} \shortrightarrow p^r$, inducing additional ``ghost'' target at the radar receiver. The ``ghost'' target would appear in the NLOS region, with reduced energy compared to the direct path and a different received echo range profile.
Although the conventional LOS multipath investigation is beyond the scope of this work, this information could be used for multipath mitigation. 

\begin{algorithm}[H]
\caption{Reflective Surface Parameters Estimation\\ (Stage I)} 
\label{algo:stage_1}
  \begin{algorithmic}[1] 
  \Statex~\textbf{Input:} $\mathbf{X}^r \in \mathbb{C}^{M_r \times N}$

  \Statex \textbf{1) Pre-Processing:}
  \State Zero padding and 2D-FFT:
 \Statex $\mathbf{Z} = \mathcal{F}_{2d} \left(P_{512}\left(\mathbf{X}^r\right)\right) \in \mathbb{C}^{512 \times 512}$
  \State Split the matrix into 3 channels:
  \Statex $\mathbf{Z}_{pre} = \left[\operatorname{Re}\{\mathbf{Z}\}, \operatorname{Im}\{\mathbf{Z}\}, |\mathbf{Z}|\right] \in \mathbb{R}^{512 \times 512 \times 3}$
  \State Downsampling and bilinear interpolation:
  \Statex $\mathbf{Z}_{pre} \rightarrow \mathbf{Z}_{in} \in \mathbb{R}^{240 \times 240 \times 3}$

  \Statex \textbf{2) CNN Processing:}
   \State $\mathbf{\zeta} = \mathbf{C}(\mathbf{Z}_{in}) \in \mathbb{R}^{1280}$

  \Statex \textbf{3) Reflective Surface Detection and Parameters Estimation:}
  \State  FC single-layer transform of each parameter:
  \Statex $\mathcal{P}^{\theta} = \sigma^{\theta}(\mathbf{W}_\theta^T \mathbf{\zeta} + \mathbf{b}_\theta): \mathbb{R}^{1280} \rightarrow \mathbb{R}^{351}$ \vspace{0.1cm}
  \Statex $\mathcal{P}^{D} = \sigma^{D}(\mathbf{W}_D^T \mathbf{\zeta} + \mathbf{b}_D): \mathbb{R}^{1280} \rightarrow \mathbb{R}^{301}$ \vspace{0.1cm}
  \Statex $\mathcal{P}^{x} = \sigma^{x}(\mathbf{W}_x^T \mathbf{\zeta} + \mathbf{b}_x): \mathbb{R}^{1280} \rightarrow \mathbb{R}^{201}$ \vspace{0.1cm}
  \Statex $\mathcal{P}^{y} = \sigma^{y}(\mathbf{W}_y^T \mathbf{\zeta} + \mathbf{b}_y): \mathbb{R}^{1280} \rightarrow \mathbb{R}^{201}$ \vspace{0.1cm}

  \State {Conditional Estimation:}
  \Statex \textbf{if} $\mathcal{P}^{\theta}_0 < \max \mathcal{P}^{\theta}_{1:350}$ \textbf{and} $\mathcal{P}^{\theta}_0 < \max \mathcal{P}^{D}_{1:300}$ \textbf{and} 
  \Statex $\mathcal{P}^{x}_0 < \max \mathcal{P}^{D}_{1:200}$ \textbf{and} $\mathcal{P}^{y}_0 < \max \mathcal{P}^{D}_{1:200}$  \textbf{then}
        \State \hspace{0.5cm} Estimate parameters of detected reflective surface:
        \Statex \hspace{0.5cm} $\hat{\theta}^{w*} = \operatorname*{arg\,max}_{\hat{\theta}^w} \mathcal{P}^{\theta}_{1:350}$ \vspace{0.1cm} 
        \Statex \hspace{0.5cm} $\hat{D}^{w*} = \operatorname*{arg\,max}_{\hat{D}^w} \mathcal{P}^{D}_{1:300}$ \vspace{0.1cm} 
        \Statex \hspace{0.5cm} $\hat{x}^{w*} = \operatorname*{arg\,max}_{\hat{x}^w} \mathcal{P}^{x}_{1:200}$ \vspace{0.1cm} 
        \Statex \hspace{0.5cm} $\hat{y}^{w*} = \operatorname*{arg\,max}_{\hat{y}^w} \mathcal{P}^{y}_{1:200}$

  \State \textbf{else}
      \State \hspace{0.5cm} No reflective surface within FOV.
  \State \textbf{end if}
    \Statex \textbf{Output:} Detected reflective surface parameters, $\hat{\mathbf{\xi}}^w \in \mathbb{R}^4$
\end{algorithmic}
\end{algorithm}

\begin{algorithm}[H]
\caption{NLOS Propagation Conditions Identification (Stage II)}
\label{algo:stage_2}
\begin{algorithmic}[1]
\Statex \textbf{Input:} Surface parameters, $\hat{\mathbf{\xi}}^w$ and $\mathbf{Z}$
\If{reflective surface was detected}
    \State create $\mathbf{M}_{NLOS}$ region mask
    \Statex \hspace{0.4cm} create $\mathbf{M}_{LOS}$ region mask
    \State \textcolor{black}{$\left\{\bar{\varphi}^*, \bar{R}^*\right\} = \operatorname*{arg\,max}_{\bar{\varphi}, \bar{R}} |\mathbf{Z} \odot (\mathbf{M}_{LOS} \lor \mathbf{M}_{NLOS})|$}
    \If{$\left\{\bar{\varphi}^*, \bar{R}^*\right\}\in \mathbf{M}_{LOS}$}
        \State decide $I_0$
    \Else
        \State decide $I_1$
    \EndIf
\Else
    \State decide $I_0$
\EndIf
\Statex \textbf{Output:} decision $I_0$ or $I_1$
\end{algorithmic}
\end{algorithm}

\subsection{Stage III: Target Parameters Estimation} \label{sec:target_e}
This subsection details the proposed approach for the target parameter estimation, summarized in Algorithm~\ref{algo:stage_3}, using the inputs:
\begin{itemize}
    \item The decision on the target propagation conditions, $I_0$ or $I_1$,
    \item Estimated parameters of the reflective surface, $\hat{\mathbf{\xi}}^w$,
    \item Estimated angle and range to the target, $\left\{\bar{\varphi}^*, \bar{R}^*\right\}$.
\end{itemize}
If the LOS propagation conditions, $I_0$, are recognized, the target location in the polar coordinates is, $\left\{\bar{\varphi}^*, \bar{R}^*\right\}$. If the NLOS propagation conditions, $I_1$, are recognized, the NLOS target location is estimated using the geometric model, introduced in Section II. The estimated range to the target, $\bar{R}^*$, is the sum of ranges from the radar to the reflective surface point, $p^w_{K_o}$, and from this point to the target at, $p^{\text{\textit{target}}}$, as:
\begin{align}
    \bar{R}^* &= \hat{R}_{p^r, p^w_{K_o}} + \hat{R}_{p^w_{K_o}, p^{\text{\textit{target}}}}\;.\label{Rfull}
\end{align}
The range from the radar to $K_o$ is estimated as:
\begin{equation} \label{eq:R_{p^r, p^w_{K_o}}}
    \hat{R}_{p^r, p^w_{K_o}} = \frac{\hat{b}^w}{\cos(\hat{\varphi}_{K_o}) - \tan(\hat{\theta}^w) \sin(\hat{\varphi}_{K_o})}\;,
\end{equation}
and the angle, $\hat{\varphi}_{K_o} = \bar{\varphi}^*$, is the estimated angle to $p^w_{K_o}$ \textcolor{black}{, obtained via digital beamforming as the peak angle in the \gls{RA} map in Eq.~\eqref{eq:argmax_Z}}. In the Cartesian coordinates, the NLOS target location is estimated as:
\begin{align} \label{eq:NLOS_target_est}
    \hat{x}^{\text{\textit{target}}} &= \hat{R}_{p^r, p^w_{K_o}} \sin (\hat{\varphi}_{K_o}) + \hat{R}_{p^w_{K_o}, p^{\text{\textit{target}}}} \sin (2\hat{\theta}^w + \hat{\varphi}_{K_o})\;, \nonumber\\
    \hat{y}^{\text{\textit{target}}} &= \hat{R}_{p^r, p^w_{K_o}} \cos (\hat{\varphi}_{K_o}) - \hat{R}_{p^w_{K_o}, p^{\text{\textit{target}}}} \cos (2 \hat{\theta}^w + \hat{\varphi}_{K_o})\;.
\end{align}

\begin{algorithm}[H]
\caption{Target Parameters Estimation (Stage III)}
\label{algo:stage_3}
\begin{algorithmic}[1]
\Statex \textbf{Input:} Decision: $I_0$ or $I_1$ , $\hat{\mathbf{\xi}}^w$, and $\left\{\bar{\varphi}^*, \bar{R}^*\right\}$
\If{$I_0$ (LOS target)}
    \State Target location is given by:
 \hspace{0.4cm} $\left\{\bar{\varphi}^*, \bar{R}^*\right\}$ in polar or $(\hat{x}^{\text{\textit{target}}}, \hat{y}^{\text{\textit{target}}})$ in Cartesian coordinates.
\Else{ (NLOS target)}
    \Statex \hspace{0.4cm} $\bar{R}^* = \hat{R}_{p^r, p^w_{K_o}} + \hat{R}_{p^w_{K_o}, p^{\text{\textit{target}}}}$ \vspace{0.2cm}
    \State Estimate range from radar to $p^w_{K_o}$:
    \Statex \hspace{0.4cm} $\hat{R}_{p^r, p^w_{K_o}} = \frac{\hat{b}^w}{\cos(\hat{\varphi}_{K_o}) - \tan(\hat{\theta}^w) \sin(\hat{\varphi}_{K_o})}$ \vspace{0.2cm}
    \State  Estimate target location: 
     \Statex \hspace{0.4cm} $\hat{x}^{\text{\textit{target}}} = \hat{R}_{p^r, p^w_{K_o}} \sin (\hat{\varphi}_{K_o}) + \hat{R}_{p^w_{K_o}, p^{\text{\textit{target}}}} \sin (2\hat{\theta}^w + \hat{\varphi}_{K_o})$
    \Statex \hspace{0.4cm} $\hat{y}^{\text{\textit{target}}} = \hat{R}_{p^r, p^w_{K_o}} \cos (\hat{\varphi}_{K_o}) - \hat{R}_{p^w_{K_o}, p^{\text{\textit{target}}}} \cos (2 \hat{\theta}^w + \hat{\varphi}_{K_o})$
\EndIf
\Statex \textbf{Output:} Target location $(\hat{x}^{\text{\textit{target}}}, \hat{y}^{\text{\textit{target}}})$
\end{algorithmic}
\end{algorithm}

\section{Performance Evaluation}
The proposed approach can be used for any type of \gls{MIMO} radar. For clarity of the following simulations, the proposed approach's performance is evaluated considering \gls{FMCW} \gls{SIMO} automotive radar in scenarios with parameters in Table~\ref{table:train_set}. The performance is evaluated in a challenging single-chirp scenario without exploiting the target's Doppler information. The received radar echo from the \gls{NLOS} target in~\eqref{eq:NLOS_target}, can be rewritten as:
\begin{align} \label{eq:RxSig_sim}
    [\mathbf{X}^{\text{\textit{target}}}]_{n} &= \Tilde{\alpha}
    \sum_{{\Tilde{k}}=1}^{\Tilde{K}} \sigma^w_{\Tilde{k}} \gamma_{\Tilde{k}} \mathbf{a}_r(\varphi_{\Tilde{k}})
    \sum_{{k'}=1}^{K'} \sigma^w_{k'} \gamma_{k'} 
    \nonumber \\
     &\cdot e^{j \pi a (\tau_{r,k'} + \tau_{k',t} + \tau_{t, \Tilde{k}} + \tau_{\Tilde{k}, r})^2} \nonumber \\
     &\cdot e^{-j 2 \pi a  (\tau_{r,k'} + \tau_{k',t} + \tau_{t, \Tilde{k}} + \tau_{\Tilde{k}, r})  \frac{T_0}{N} n }\;,
\end{align}
where the linear \gls{FMCW} transmit waveform is $s_t(t) = e^{j\pi a t^2}$, where \textcolor{black}{$a=\frac{BW}{T_0}$} is the chirp slope, where \textcolor{black}{$BW$} is the radar bandwidth and $T_0$ is the chirp duration.

The performance of the proposed \gls{NLOS} target localization approach was evaluated using a simulated dataset of $6.4$ million frames of received radar echoes in NLOS scenarios, generated using the model, detailed in Section \ref{sec:problem_def}. The parameters of the considered scenarios were simulated according to Table~\ref{table:train_set}. 
The appearance of the reflective surface in the training dataset was randomized in the following $4$ scenarios:
\begin{enumerate}
    \item \gls{NLOS} target.
    \item \gls{LOS} target:
    \begin{itemize}
        \item without a reflective surface in the scene.
        \item with a reflective surface in the scene but without multipath. 
        \item with a reflective surface in the scene and multipath. 
\end{itemize}
\end{enumerate}

\textcolor{black}{Since the data is synthetically generated, splitting it into training and testing datasets is unnecessary. Instead, new unseen data, independent from the training and testing datasets, is generated for the validation dataset. In addition, no pretraining was used, and all models were trained from scratch.}
The $R_{p^r, p^w_{K_o}}$ were estimated using \eqref{eq:R_{p^r, p^w_{K_o}}}, and the target location in Cartesian coordinates was estimated using \eqref{eq:NLOS_target_est}. 

All the simulations were performed using AMD Ryzen Threadripper PRO 5965WX, with an Nvidia RTX A5000 Ada GPU. An ADAM optimizer~\cite{kingma2017adammethodstochasticoptimization} with a polynomial decay policy~\cite{bukhari2021systematic} to control the learning rate, was considered. The optimization was initiated with a learning rate of $10^{-4}$, a decaying power of $0.9$, and terminated with a learning rate of $10^{-5}$. The training was performed using a batch size of $32$.

\begin{table}[!ht]  
\begin{center}
\caption{Training dataset simulation parameters.}
 \begin{tabular}{m{1.2cm} m{3.7cm} m{2.7cm}}
 \hline\hline
 Notation & Description & Value \\ [1.0ex] 
 \hline\hline
 $x^w$ & Reflective surface center x & $\sim\mathcal{U}\left(\left[0m,6m\right]\right)$ \\ [1.0ex]
 \hline
 $y^w$ & Reflective surface center y & $\sim\mathcal{U}\left(\left[8m,22m\right]\right)$ \\ [1.0ex]
 \hline
 $\theta^w$ & Reflective surface orientation angle & $\sim\mathcal{U}\left([1^\circ,46^\circ]\right)$ \\ [1.0ex]
 \hline
  $D^w$ & Reflective surface length & $\sim\mathcal{U}\left([1m,13m]\right)$ \\ [1.0ex]
 \hline
  ${R}_{p^w_{K_o}, p^{\text{\textit{target}}}}$ & Reflective surface to target range & $\sim\mathcal{U}\left([6m,11m]\right)$ \\ [1.0ex]
 \hline
  $\varphi_{K_o}$ & Angle to $p^w_{K_o}$& $\sim\mathcal{U}\left(\varphi_{p^w_1},\varphi_{p^w_L}\right)\;[^\circ]$ \\ [1.0ex]
 \hline
 $M_t$ & Number of transmitters & $M_t=1$ \\ [1.0ex]
 \hline
   $M_r$ & Number of receivers & $M_r=16$ \\ [1.0ex]
 \hline
  $N$ & Fast-time samples & $N=128$ \\ [1.0ex]
 \hline
   \textcolor{black}{$BW$} & Band-width & \textcolor{black}{$BW = 400~\mathrm{MHz}$}
 \\ [1.0ex]
 \hline
 $\text{$\text{SNR}^{\text{\textit{target}}}$}$ & Target SNR & $\sim\mathcal{U}\left([0,80]\right)\;[\,\mathrm{dB}]$ \\ [1.0ex]
 \hline
 $\text{$\text{SNR}^w$}$ & Reflective surface SNR & $\sim\mathcal{U}\left([0,70]\right)\;[\,\mathrm{dB}]$ \\ [1.0ex]
 \hline
  $\Lambda$ & Repartition factor & $\Lambda=0.846$ \\ [1.0ex]
 \hline
  $\psi^w$ & Beamwidth & $\psi^w=14$ \\ [1.0ex]
 \hline
 \label{table:train_set}
\end{tabular}
\end{center}
\end{table}

First, Subsection A evaluates the performance of the reflective surface parameters, $\hat{\mathbf{\xi}}^w$, estimation. Next, the \gls{NLOS} propagation conditions identification performance is evaluated in Subsection B. Finally, the performance of the target location estimation, $(\hat{x}^{\text{\textit{target}}}, \hat{y}^{\text{\textit{target}}})$, is evaluated in Subsection C. 
\textcolor{black}{The post-processing \glspl*{SNR} are defined after matched filtering and coherent integration across $M_t$ transmitters, $M_r$ receivers, and $N$ receiver samples. The SNR of the echo from the LOS reflective surface is defined as:}
\begin{equation} \label{eq:surface_SNR}
    \text{SNR}^w = \frac{|\alpha_k^w|^2}{\sigma_n^2} + 10 \log_{10}(M_t M_r N) \;,
\end{equation}
\textcolor{black}{and the \gls*{SNR} of the radar echo from the target is defined as:}
\begin{equation} \label{eq:target_SNR}
    \text{SNR}^{\text{\textit{target}}} = \frac{|\sigma^{\text{\textit{target}}}|^2}{\sigma_n^2} + 10 \log_{10}(M_t M_r N) \;,
\end{equation}
where $\alpha_k^w$ is defined in \eqref{eq:wall_model}, $\sigma^{\text{\textit{target}}}$ is defined in \eqref{eq:mimo_target} and $\sigma_n^2$ is the noise variance. 
Since the strong reflection from the reflective surface can mask the weak \gls{NLOS} target, the \gls{NLOS} target localization performance is also evaluated using the differential SNR, $(\Delta \text{SNR})$, defined as:
\begin{equation} \label{eq:Detla_SNR}
    \Delta \text{SNR} \triangleq \frac{|\sigma^{\text{\textit{target}}}|^2 - |\alpha^w_k|^2}{\sigma_n^2}\;.
    \end{equation}

\subsection{Reflective Surface Parameters Estimation (Stage I)} \label{sec:wall_param_eval}
This subsection evaluates the performance of the estimation of reflective surface parameters. First, the performance of various \gls{DNN} architectures is evaluated. Next, the performance of the proposed approach with the optimal \gls{DNN} architecture is compared with the conventional estimators. Finally, the influence of the reflective surface orientation and length on the estimation performance is evaluated.
\subsubsection{DNN Architecture Selection}
First, the performance of {\it ConViT}~\cite{d2021convit}, {\it EfficientNet $b_0$}, and {\it EfficientNet $b_1$} architectures of reflective surface parameters, $\hat{\mathbf{\xi}}^w$, estimation, were evaluated via \gls{MC} simulations in scenarios with various $\text{SNR}^w$ . The reflective surface with the following parameters, ${x}^w = 2m,\; {y}^w = 18m,\; {D}^w = 8m,\; \theta^w=25^\circ$, was simulated. Fig.~\ref{fig:wall_meter} shows the performance of $[\hat{x}^w, \hat{y}^w, \hat{D}^w]$  and $\hat{\theta}^w$ estimation in scenarios with $\text{SNR}^w$s in the range between $0$ to $50\,\mathrm{dB}$. 

Fig.~\ref{fig:wall_meter} shows that the \glspl{RMSE} of the simulated estimators decrease with increasing $\text{SNR}^w$, and the {\it EfficientNet} architectures outperform the {\it ConViT}.
\textcolor{black}{This can be attributed to the fact that the \gls{RA} map exhibits strong local spatial structures, which align well with the locality-focused nature of convolutional networks like {\it EfficientNet}, whereas {\it ConViT}'s global attention mechanism may be less efficient for this type of data.}
Notice that {\it EfficientNet $b_1$} outperforms the {\it EfficientNet $b_0$} 
at low SNRs due to its larger number of learnable parameters, which enhances its modeling capability. At higher SNRs, the performance of {\it EfficientNet $b_0$} and {\it EfficientNet $b_1$} are similar. According to these results, the {\it EfficientNet $b_1$} is considered for the performance evaluations in the next subsections. 

\begin{figure}
     \centering
     \begin{subfigure}[b]{0.48\textwidth}
         \centering
         \includegraphics[width=\textwidth]{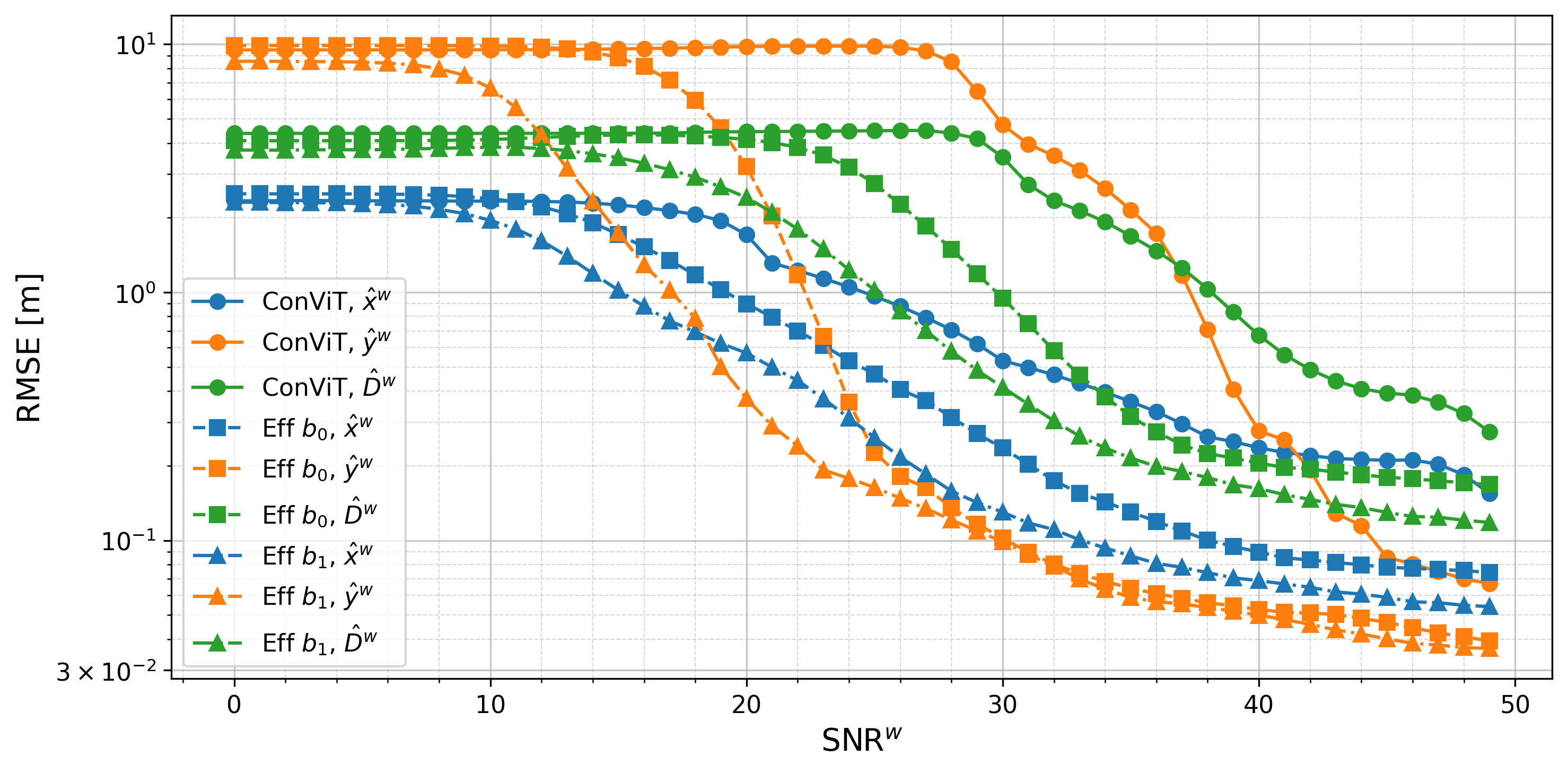}
         \caption{RMSE of $[\hat{x}^w, \hat{y}^w, \hat{D}^w]$ estimation as a function of $\text{SNR}^w$.}    
             \end{subfigure}
     \hfill 
     \begin{subfigure}[b]{0.48\textwidth}
         \centering
         \includegraphics[width=\textwidth]{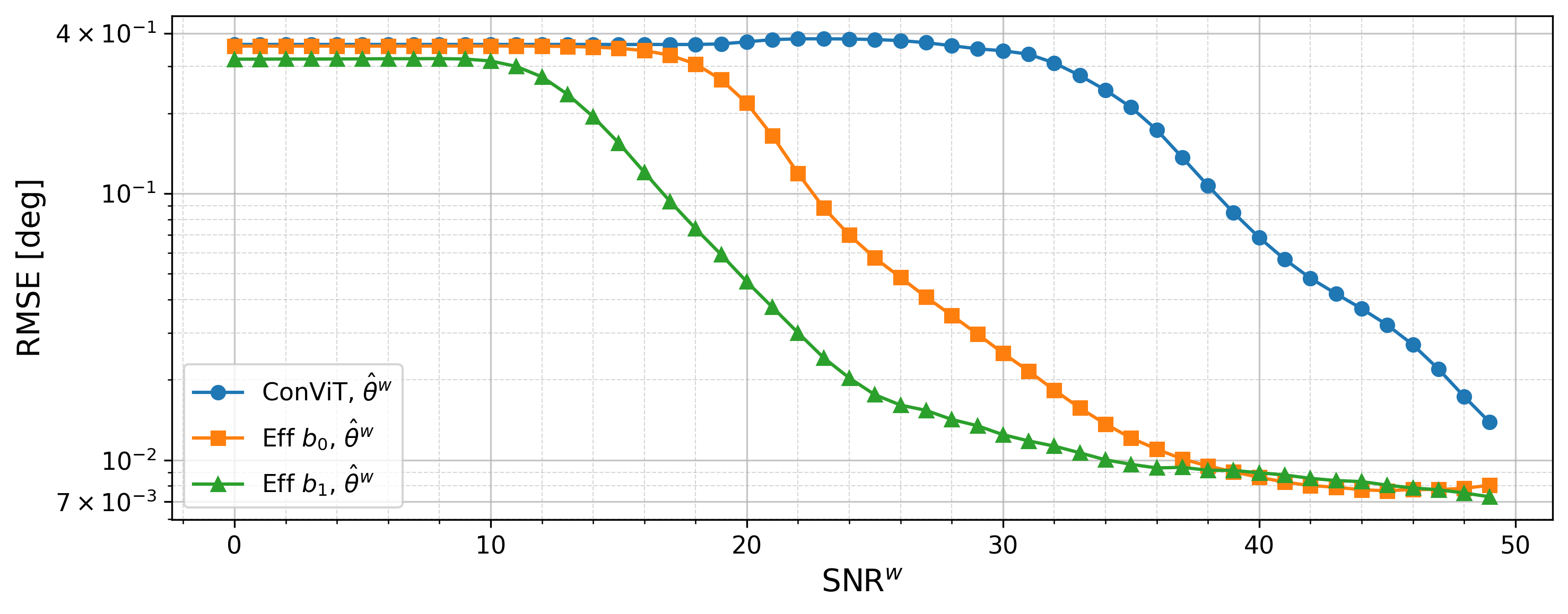}
         \caption{RMSE of $\hat{\theta}^w$ estimation as a function of $\text{SNR}^w$.}
     \end{subfigure}
     \caption{RMSE of reflective surface parameters estimation of {\it ConViT}, {\it EfficientNet $b_0$}, and {\it EfficientNet $b_1$} architectures, in scenarios with the reflective surface parameters, ${x}^w = 2m, {y}^w = 18m, {D}^w = 8m, \theta^w=25^\circ$. \textcolor{black}{Performance was evaluated using $3.8 \times 10^4$ \gls{MC} simulations.}}
      \label{fig:wall_meter}
\end{figure}
\subsubsection{Performance of the Reflective Surface Parameters Estimation}
Next, the performance of the proposed approach with {\it EfficientNet $b_1$} architecture for the reflective surface parameters estimations is compared with the conventional \gls{LS}~\cite{alma9926318476504361} and \gls{RANSAC}~\cite{fischler1981random} estimators. Notice that, unlike the proposed approach, \gls{LS} and \gls{RANSAC} that approximate the \gls{ML} estimator to identify ${k'}$s scatterers, require prior knowledge of the number of reflectors, $K$. The estimated range-DOA bins, $(\hat{\tau}_{r,{k'}}, \hat{\varphi}_{k'})$, for each ${k'}$-th reflector are converted into the Cartesian coordinates, $(\hat{x}_{k'}, \hat{y}_{k'})$.

\textcolor{black}{The input coordinates are extracted from the \gls*{RA} map, which is first generated by 2D FFT processing across the fast-time and channel dimensions of the radar signal, followed by magnitude computation, yielding $|\mathbf{Z}|$. An iterative peak detection algorithm is then applied to identify the $K$ strongest local maxima in the \gls*{RA} map, corresponding to dominant reflections in the scene. Each detected peak provides an estimated \gls*{RA} pair, $(\hat{\tau}_{r,{k'}}, \hat{\varphi}_{k'})$, which is then transformed into Cartesian coordinates based on the radar's known geometric configuration. These coordinates serve as input to the \gls*{LS} and the \gls*{RANSAC}-based surface parameter estimators.}

The \gls{LS} minimizes the following objective function:
\begin{equation}
    \hat{\mathbf{\xi}}^{w^{LS*}} = \operatorname*{arg\,min}_{\hat{\mathbf{\xi}}^{w^{LS}}} \|\hat{\mathbf{y}} - \hat{\mathbf{H}} \hat{\mathbf{\xi}}^{w^{LS}}\|^2\;,
\end{equation}
which for the linear model is:
\begin{equation}
    \hat{\mathbf{\xi}}^{w^{LS*}} = (\hat{\mathbf{H}}^T \hat{\mathbf{H}})^{-1} \hat{\mathbf{H}}^T \hat{\mathbf{y}}\;,
\end{equation}
where $\hat{\mathbf{y}} = [\hat{y}_1, ..., \hat{y}_{k'}]^T$ and $\hat{\mathbf{H}} = [\mathbf{1}_{k'}, [\hat{x}_1, ..., \hat{x}_{k'}]^T]$. The \gls{RANSAC} fits a linear model by splitting the data into inliers and outliers and computes the parameters using only the inliers, enhancing robustness against outliers.

Fig.~\ref{fig:wall_meter_eff_ls_ransac} shows the \gls{RMSE} of the reflective surface parameters estimated as a function of $\text{SNR}^w$, of {\it EfficientNet $b_1$}, \gls{LS}, and \gls{RANSAC}.
The proposed {\it EfficientNet $b_1$} outperforms the \gls{LS}, and \gls{RANSAC}, especially at higher SNRs. This stems from the fact that the proposed approach does not require {\it a-priori} knowledge on $K$ and is not limited by the resolution constraints, typically associated with point-cloud \gls{RA} output from the \gls{ML}-like estimators. The proposed approach, whose resolution is determined by the dimensions of the considered \gls{DNN} architecture only, resolves this limitation of the ML-like estimators, enabling more detailed and accurate surface parameter reconstructions.

\begin{figure}
     \centering
     \begin{subfigure}[b]{0.48\textwidth}
         \centering
         \includegraphics[width=\textwidth]{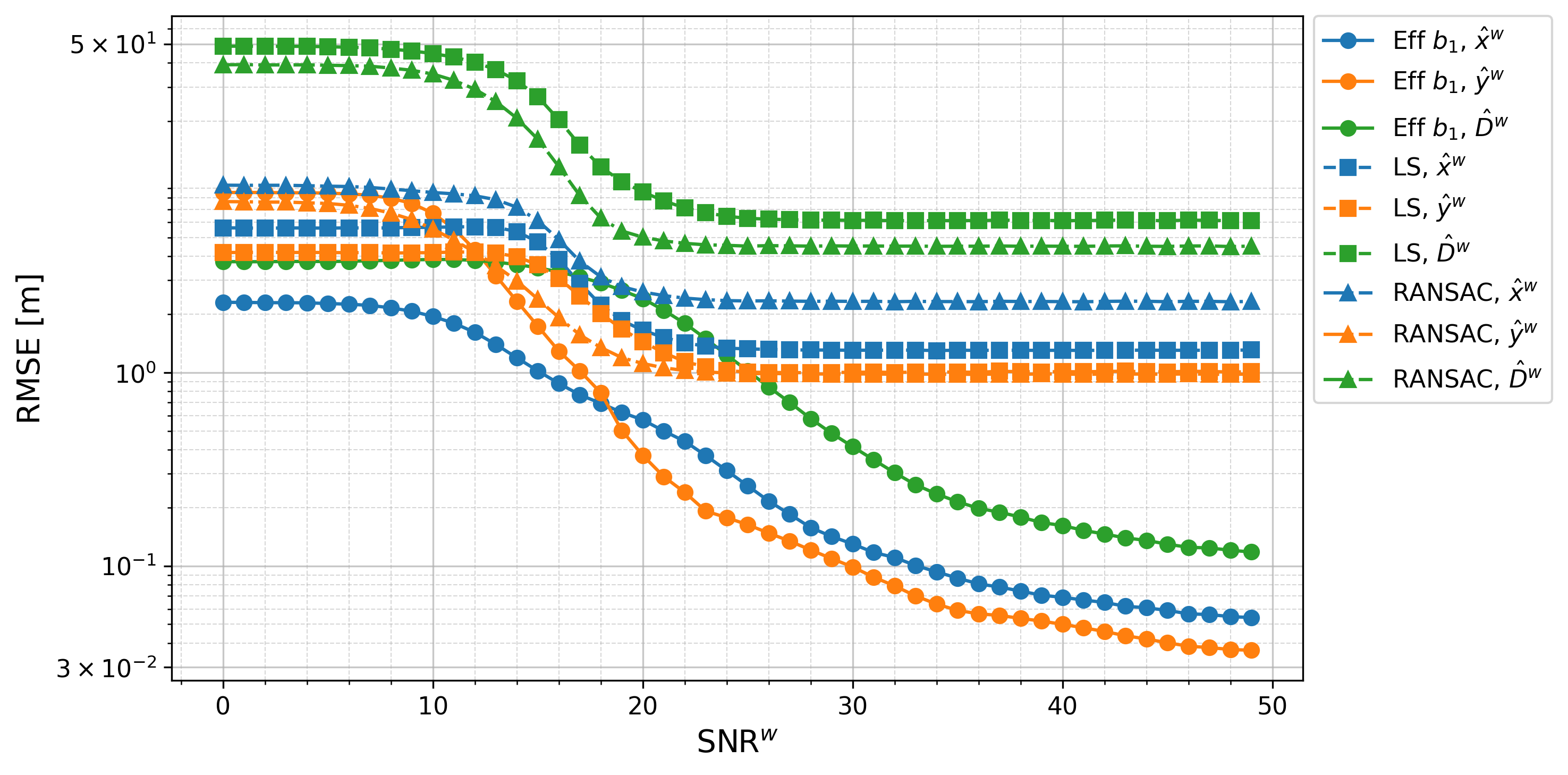}
         \caption{RMSE of $[\hat{x}^w, \hat{y}^w, \hat{D}^w]$ estimation as a function of $\text{SNR}^w$.}    
     \end{subfigure}
     \hfill 
     \begin{subfigure}[b]{0.48\textwidth}
         \centering
         \includegraphics[width=\textwidth]{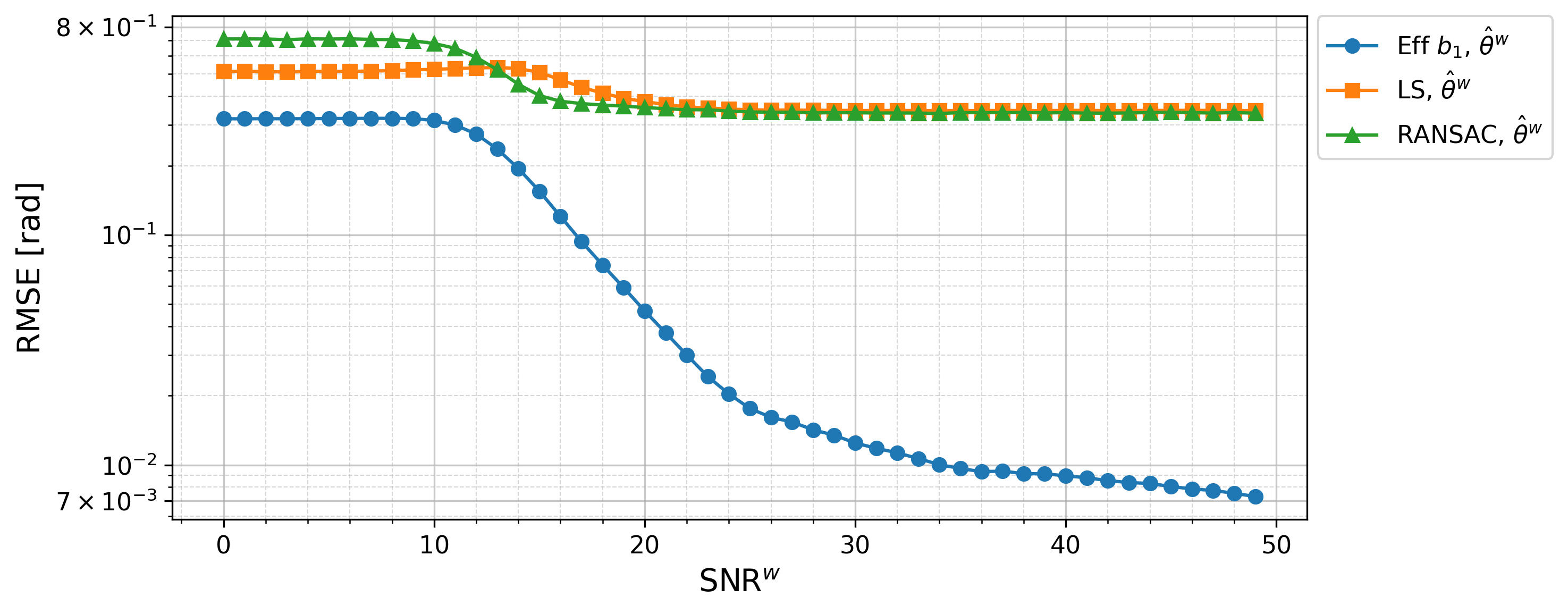}
         \caption{RMSE of $\hat{\theta}^w$ estimation as a function of $\text{SNR}^w$.}
     \end{subfigure}
     \caption{RMSE of reflective surface parameters estimation of {\it EfficientNet $b_1$}, \gls{LS} and RANSAC in scenarios with the reflective surface parameters, ${x}^w = 2m, {y}^w = 18m, {D}^w = 8m, \theta^w=25^\circ$. \textcolor{black}{Performance was evaluated using $2.2 \times 10^4$ \gls{MC} simulations.}}
      \label{fig:wall_meter_eff_ls_ransac}
\end{figure}

\subsubsection{Influence of Reflective Surface Orientation} \label{sec:wall_param_angle_eval}
This subsection evaluates the influence of the reflective surface orientation, $\theta^w$, on the performance of the parameters,  $\hat{\mathbf{\xi}}^w$, estimation. Fig.~\ref{fig:wall_param_angle_eval} shows the \gls{RMSE} of the reflective surface parameters estimation as a function of its orientation in scenarios with the $\text{SNR}^w=[20, 30, 40]\,\mathrm{dB}$. Notice that the estimation errors of all parameters decrease with increasing $\theta^w$. 
This observation can be explained by the increasing number of observable reflectors within each range-DOA bin, with increasing angle of the reflective surface orientation, $\theta^w$.

\begin{figure}
     \centering
     \begin{subfigure}[b]{0.48\textwidth}
         \centering
         \includegraphics[width=\textwidth]{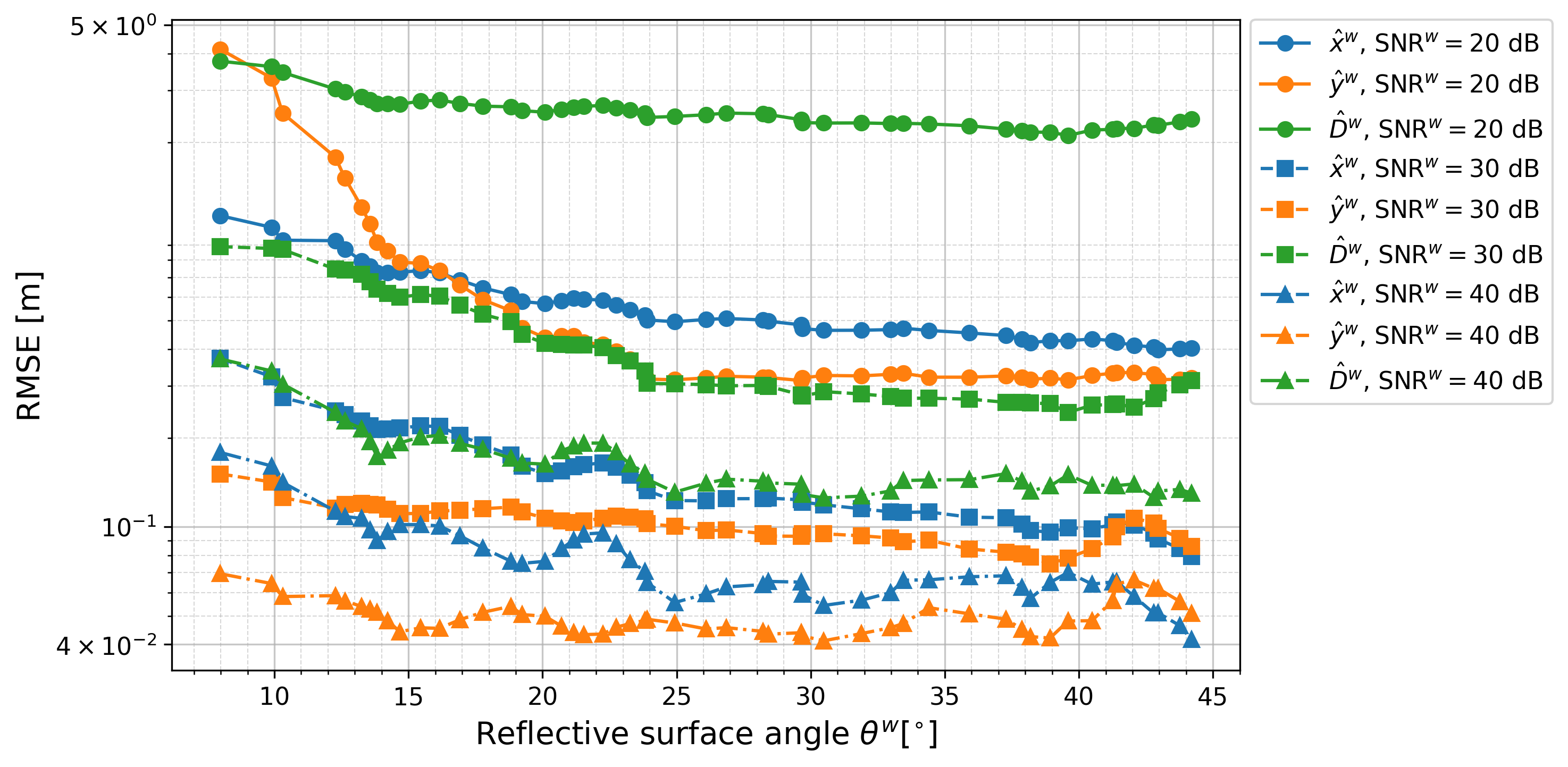}
        \caption{RMSE of $[\hat{x}^w, \hat{y}^w, \hat{D}^w]$ as a function of reflective surface orientation, $\theta^w$.}
         \label{fig:wall_x,y,length,_wall_angle_eval}
     \end{subfigure}
     \hfill 
     \begin{subfigure}[b]{0.48\textwidth}
         \centering
         \includegraphics[width=\textwidth]{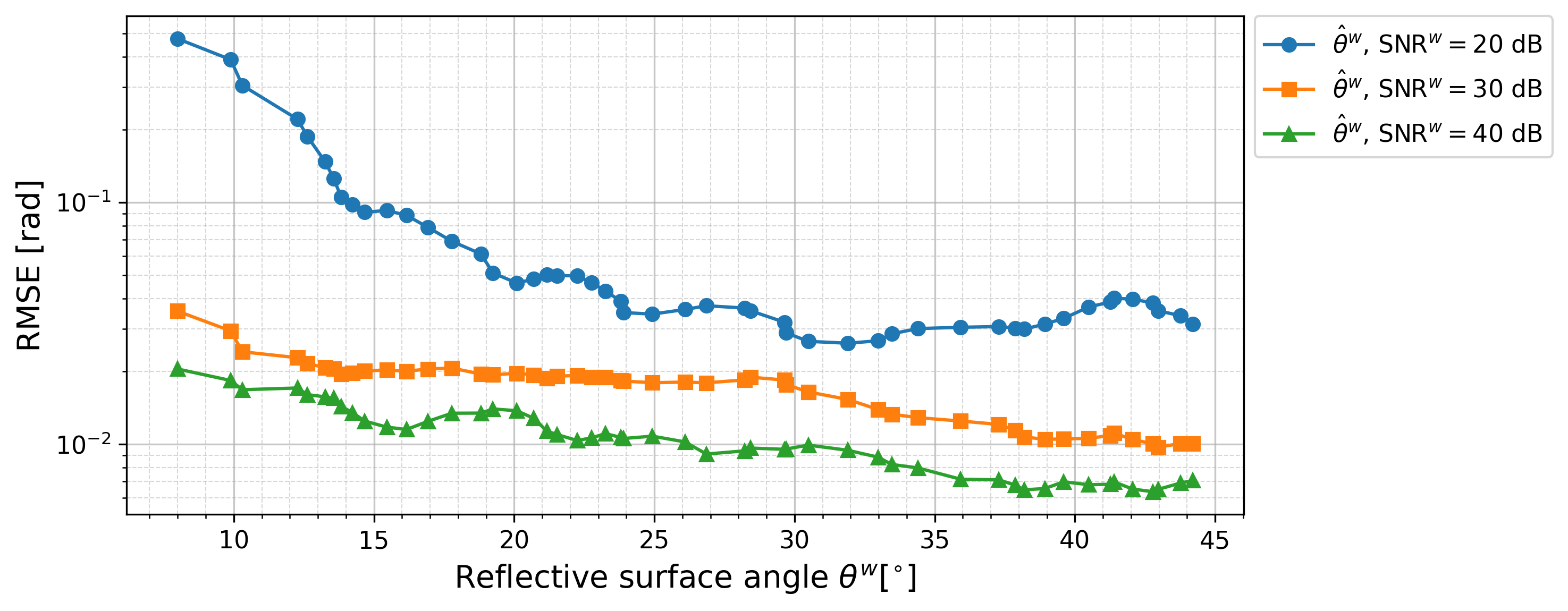}
        \caption{RMSE of $\hat{\theta}^w$ as a function of reflective surface orientation, $\theta^w$.}
         \label{fig:wall_angle_wall_angle_eval}
     \end{subfigure}
    \caption{\gls{RMSE} of reflective surface parameters estimation as a function of its orientation, $\theta^w=[5^\circ:45^\circ]$, in scenarios with the parameters, ${x}^w = 3m, {y}^w = 18m, {D}^w = 8m$, and $\text{SNR}^w=[20, 30, 40]\,\mathrm{dB}$. \textcolor{black}{Performance was evaluated using $2.9 \times 10^4$ \gls{MC} simulations.}
    } 
    \label{fig:wall_param_angle_eval}
\end{figure}

\subsubsection{Influence of Reflective Surface Length} \label{sec:wall_param_len_eval}
This subsection evaluates the influence of the reflective surface length, $D^w$, on its parameter estimation performance. Fig.~\ref{fig:wall_len_eval} shows that the RMSE of the reflective surface parameters estimation improves with increasing length, $D^w$.
This observation can be explained by the increasing number of reflectors on the reflective surface that contribute to the estimation of its parameters. 

\begin{figure}
     \centering
     \begin{subfigure}[b]{0.48\textwidth}
         \centering
         \includegraphics[width=\textwidth]{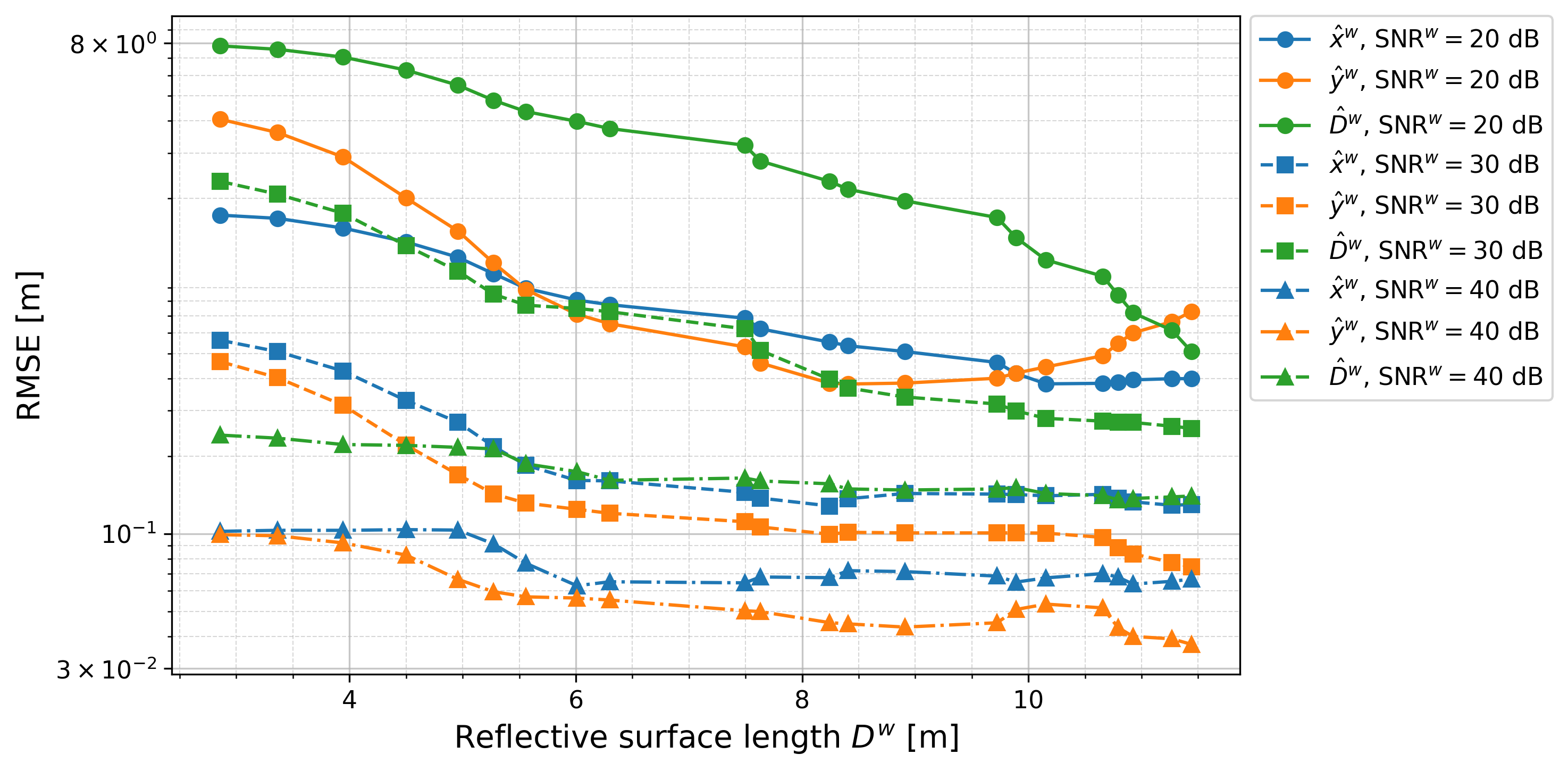}
         \caption{RMSE of $[\hat{x}^w, \hat{y}^w, \hat{D}^w]$ estimation as a function of $D^w$.}    
         \label{fig:wall_x,y,length,_wall_len_eval}
     \end{subfigure}
     \hfill 
     \begin{subfigure}[b]{0.48\textwidth}
         \centering
         \includegraphics[width=\textwidth]{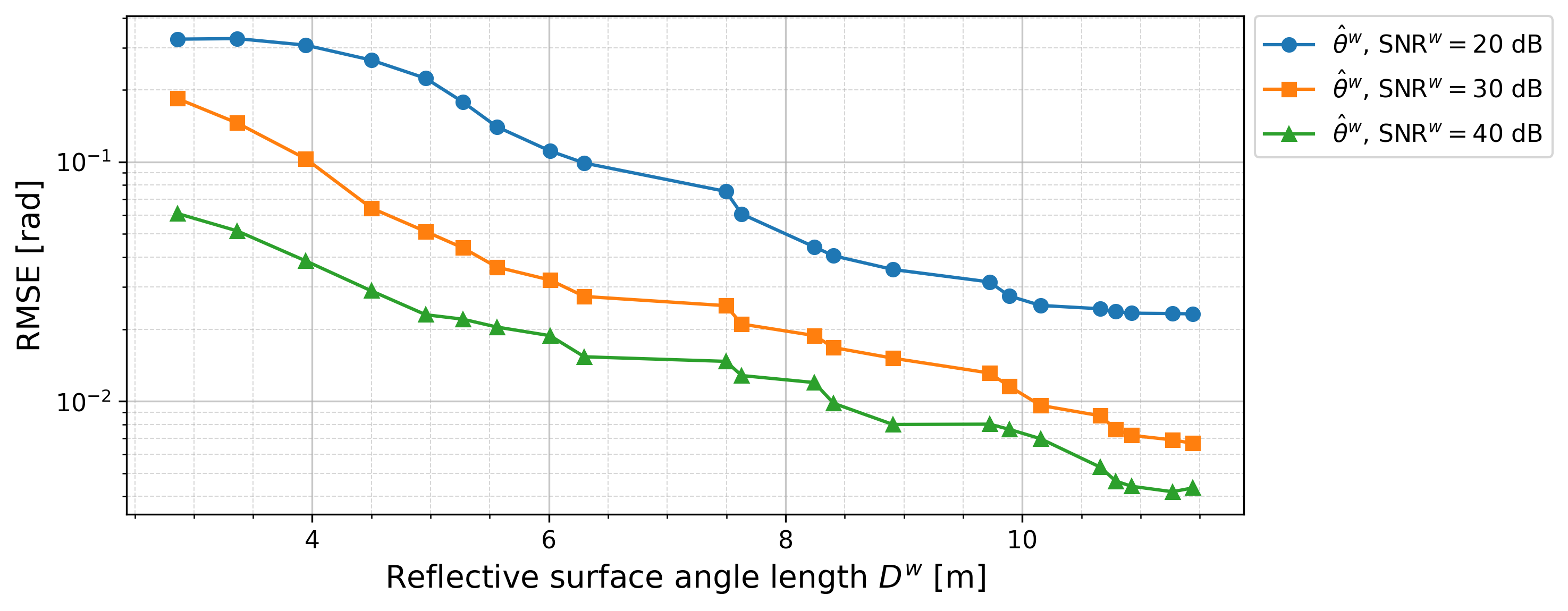}
         \caption{RMSE of $\hat{\theta}^w$ estimation as a function of $D^w$.}
         \label{fig:wall_angle_wall_len_eval}
     \end{subfigure}
\caption{\gls{RMSE} of reflective surface parameters estimation as a function of its length, $D^w=[1:11]m$ in scenarios with the reflective surface parameters, ${x}^w = 2m, {y}^w = 18m, \theta^w = 25^\circ$, and $\text{SNR}^w=[20, 30, 40]\,\mathrm{dB}$. \textcolor{black}{Performance was evaluated using $4.6 \times 10^4$ \gls{MC} simulations.} 
}
\label{fig:wall_len_eval}
\end{figure}

\subsection{NLOS Propagation Conditions Identification (Stage II)} \label{sec:NLOS_LOS_eval}
This subsection evaluates the performance of the LOS/NLOS propagation conditions identification considering two scenarios: 1) $I_1$ - NLOS propagation conditions and 2) $I_0$ - LOS with or without reflective surface in the scene. In both conditions, the reflective surface is simulated at randomized locations with the parameters, generated from a uniform distribution, $\mathcal{U}(a, b)$, where $a$ and $b$ are the lower and upper bounds of the distribution: $x^w \sim \mathcal{U}[0\,\text{m},\; 6\,\text{m}]$, $y^w \sim \mathcal{U}[12\,\text{m},\; 22\,\text{m}]$, $D^w \sim \mathcal{U}[4\,\text{m},\; 12\,\text{m}]$, $\theta^w \sim \mathcal{U}[1^\circ,\; 45^\circ]$. In the $I_1$ scenarios, the NLOS target is randomized with the following parameters: $\varphi_{K_o} \sim \mathcal{U}\left[( 1.25 \varphi_{p^w_1})^\circ, (0.75 \varphi_{p^w_L})^\circ \right]$, $R_{p^w_{K_o}, p^{\text{\textit{target}}}} \sim \mathcal{U}[7\,\text{m}, 18\,\text{m}]$. In the $I_0$ scenarios with the reflective surface in the scene, the LOS target location is randomized within the NLOS area. In scenarios with $I_0$ conditions and without the reflective surface in the scene, the LOS target location is randomized within the entire radar FOV.

Fig.~\ref{fig:pd_pfa_delta_SNR} shows the probability of correctly identifying the NLOS propagation conditions, $\Pr(I_1 | I_1)$, and the false alarms, $\Pr(I_1 | I_0)$, as a function of $\Delta \text{SNR}$.
Notice that at $\text{SNR}^w =30\,\mathrm{dB}$, the NLOS propagation conditions are correctly identified by all considered DNN architectures. At low $\text{SNR}^w$, the {\it EfficientNet $b_1$} outperforms the other simulated architectures and approaches $\Pr(I_1 | I_1)=1$ at $\Delta \text{SNR}>30\,\mathrm{dB}$.

\begin{figure}
     \centering
     \begin{subfigure}[b]{0.48\textwidth}
         \centering
         \includegraphics[width=\textwidth]{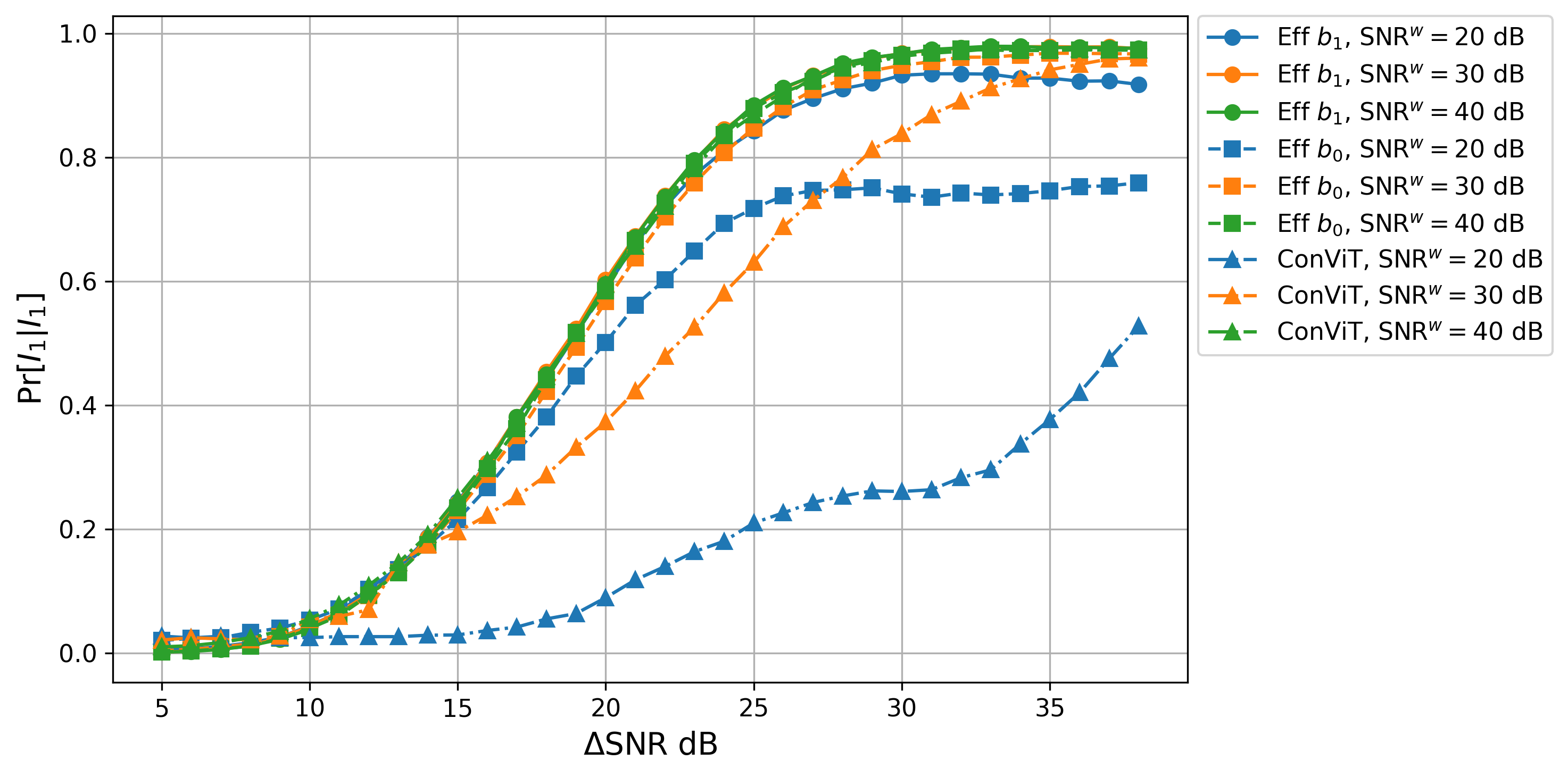}
         \caption{$\Pr(I_1 | I_1)$ as a function of the $\Delta \text{SNR}$.}    
         \label{fig:pd_delta_SNR}
     \end{subfigure}
     \hfill 
     \begin{subfigure}[b]{0.48\textwidth}
         \centering
         \includegraphics[width=\textwidth]{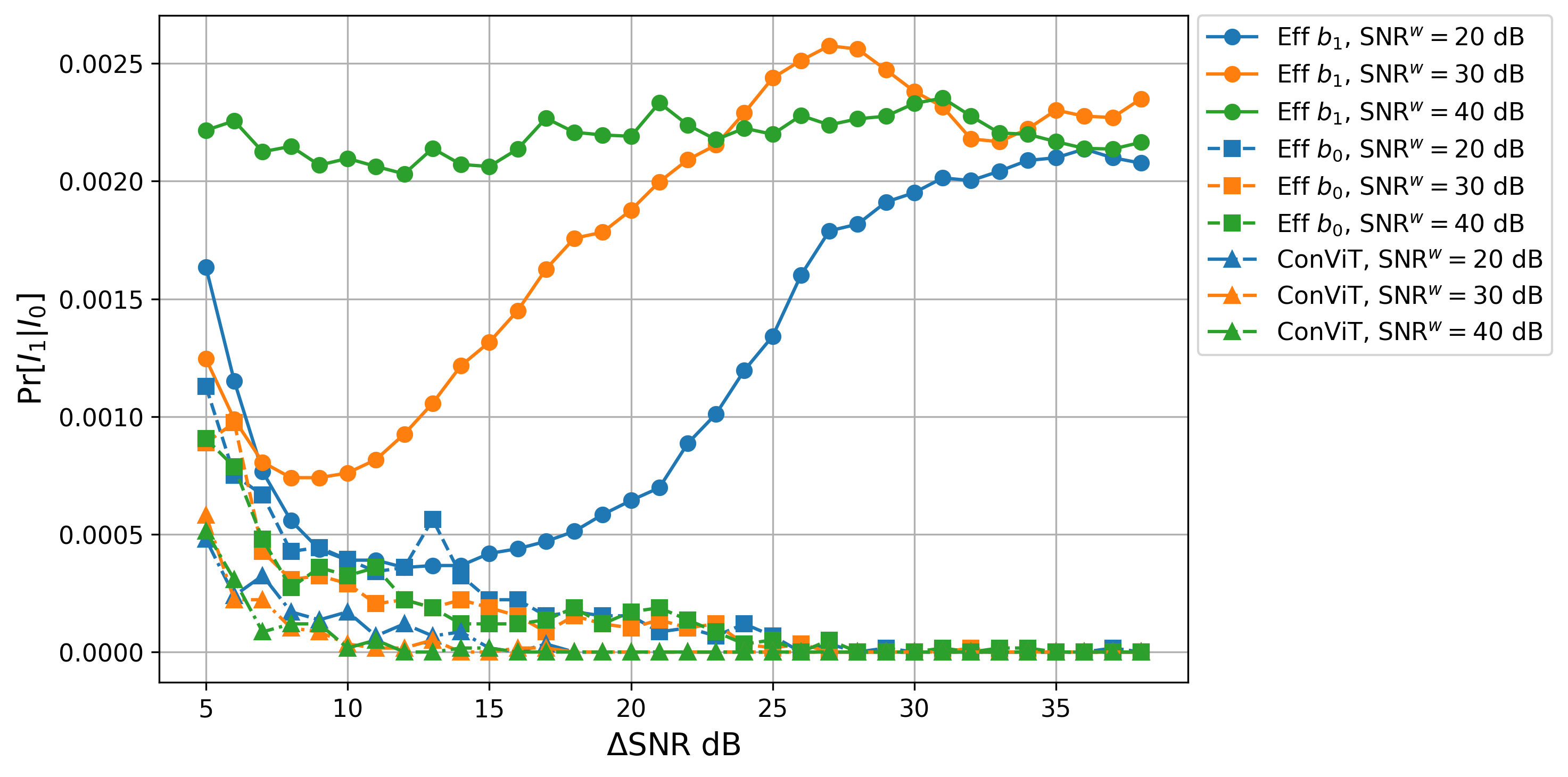}
         \caption{$\Pr(I_1 | I_0)$ as a function of the $\Delta \text{SNR}$.}
         \label{fig:pfa_delta_SNR}
     \end{subfigure}
\caption{NLOS propagation conditions identification using {\it ConViT}, {\it EfficientNet $b_0$}, and {\it EfficientNet $b_1$} architectures, as a function of the $\Delta \text{SNR}$ in scenarios with the reflective surface parameters, $x^w \sim \mathcal{U}[0\,\text{m}, 6\,\text{m}]$, $y^w \sim \mathcal{U}[12\,\text{m}, 22\,\text{m}]$, $D^w \sim \mathcal{U}[4\,\text{m}, 12\,\text{m}]$, $\theta^w \sim \mathcal{U}[1^\circ, 45^\circ]$, target parameters: $\varphi_{K_o} \sim \mathcal{U}\left[ (1.25 \varphi_{p^w_1})^\circ, (0.75 \varphi_{p^w_L})^\circ \right]$, $R_{p^w_{K_o}, p^{\text{\textit{target}}}} \sim \mathcal{U}[7\,\text{m}, 18\,\text{m}]$, and $\text{SNR}^w=[20, 30, 40]\,\mathrm{dB}$. \textcolor{black}{Performance was evaluated using $5.8 \times 10^4$ \gls{MC} simulations.}}
    \label{fig:pd_pfa_delta_SNR}
\end{figure}

\subsection{NLOS Target Localization (Stage III)} \label{sec:target_eval}
This subsection evaluates the NLOS target localization performance using two criteria. First, the proposed approach evaluates the RMSE of the range, $\bar{R}^*$ from~\eqref{Rfull}, and the direction, $\hat{\varphi}_{K_o}$, estimation, without incorporating information on the reflective surface. Next, combining these results with the reflective surface parameters estimation performance from Subsection A, the NLOS target location, $(\hat{x}^{\text{\textit{target}}}, \hat{y}^{\text{\textit{target}}})$, estimation performance is evaluated using the following criterion in the Euclidean space: 
\begin{equation}
    \mathrm{RMSE}_d = \sqrt{\mathbf{E}[(\hat{x}^{\text{\textit{target}}}-x^{\text{\textit{target}}})^2 + (\hat{y}^{\text{\textit{target}}}-y^{\text{\textit{target}}})^2]}\;.
\end{equation}

\textcolor{black}{The following experiments show that the low $\mathrm{RMSE}_d$ is achieved at relatively high post-processed \glspl*{SNR}, compared with the conventional \gls*{LOS} scenarios, due to the additional interaction of propagating waves with the reflective surface. However, these \glspl*{SNR} are practically achievable at the automotive radars with wide BW and a large number of antenna elements in the short-range urban scenarios with shorter propagation paths and a larger illuminated reflective surface, which induce a higher reflection gain of the considered diffusive model~\cite{4052607}. Notice that the short-range scenarios are those where the proposed \gls*{NLOS} targets localization approach is the most valuable due to the limited response time to the potential threat.}

\subsubsection{\texorpdfstring{$\mathrm{RMSE}_d$ as Function of $\Delta$SNR}{RMSE as Function of Delta SNR}}
Fig.~\ref{fig:target_radios_rmse_delta_SNR} shows the $\mathrm{RMSE}_d$ of the NLOS target localization as a function of the $\Delta$\gls{SNR} for the simulated $\text{SNR}^w$ of $[10:50]\,\mathrm{dB}$.
Notice that increasing the $\text{SNR}^w$ enables lower $\mathrm{RMSE}_d$. 
In scenarios with low $\text{SNR}^w$, below $10\,\mathrm{dB}$, the \gls{RMSE} is high for all simulated $\Delta$\gls{SNR} values. Notice that at high $\Delta$\gls{SNR}, beyond $25\,\mathrm{dB}$, the proposed approach can achieve a high accuracy of \gls{NLOS} target localization.

\begin{figure}[htp]
    \centering 
    \includegraphics[width=0.48\textwidth]{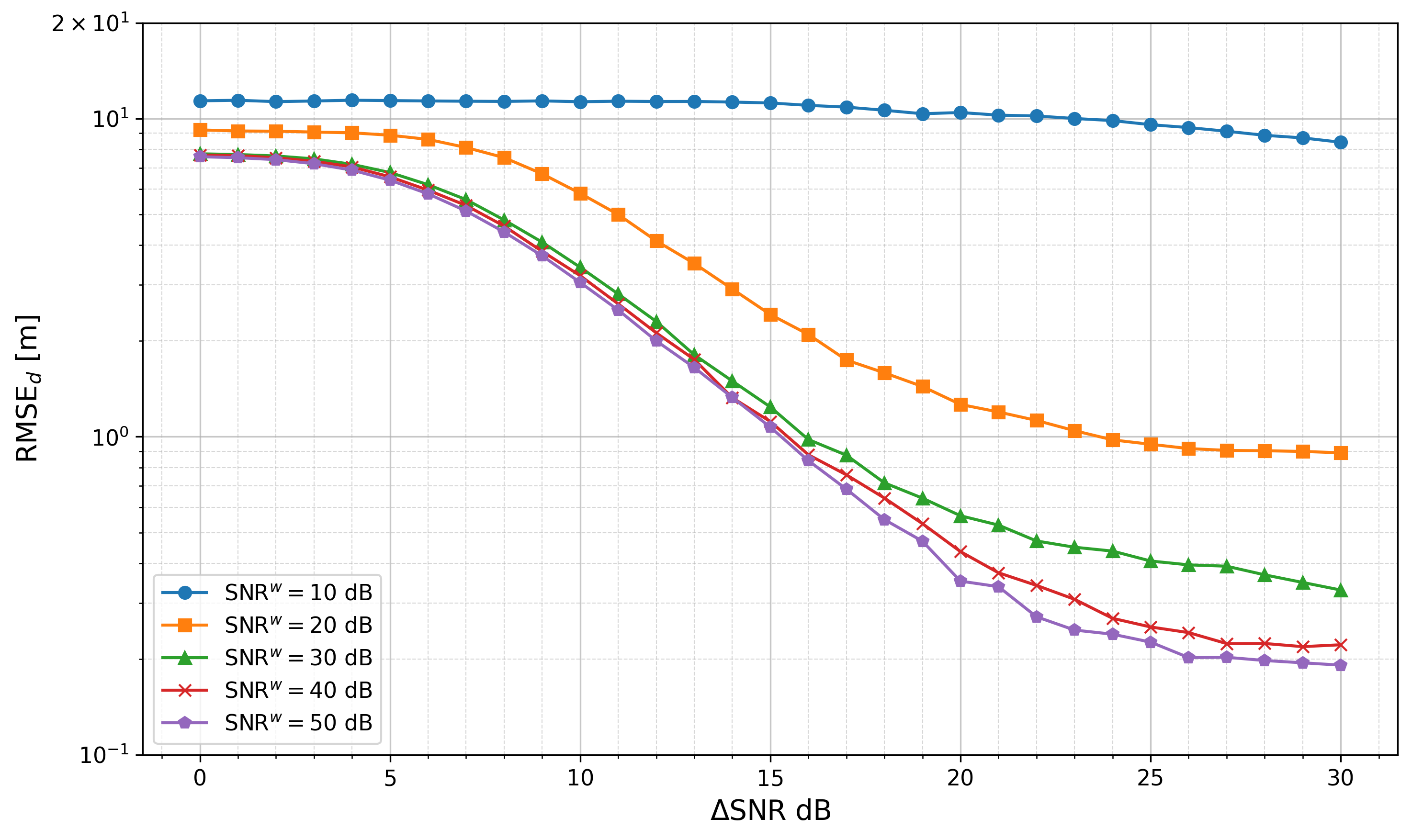}
    \caption{$\mathrm{RMSE}_d$ as function of $\Delta$\gls{SNR} in scenarios with the reflective surface parameters,  $\text{SNR}^w=[10, 20, 30, 40, 50]\,\mathrm{dB}$, ${x}^w = 2m, {y}^w = 18m, {D}^w = 8m, \theta^w=25^\circ$ and the target parameters, $\varphi_{K_o}=6.3^\circ, R_{p^r, p^w_{K_o}} = 18.1m, {R}_{p^w_{K_o}, p^{\text{\textit{target}}}} = 11.9m$. \textcolor{black}{Performance was evaluated using $2.2 \times 10^4$ \gls{MC} simulations.}}
    \label{fig:target_radios_rmse_delta_SNR}
\end{figure}

\subsubsection{\texorpdfstring{Influence of the Reflective Surface Orientation, $\theta^w$}{Influence of the Reflective Surface Orientation, theta w}}
\label{sec:eval_wall_ori}
Fig.~\ref{fig:target_rmse_wall_angle_eval} shows that the RMSE of the NLOS target parameters,  $\bar{R}^*$ and $\hat{\varphi}_{K_o}$ estimation degrades with increasing reflective surface orientation, $\theta^w$. 
This can be explained by the observation that the number of reflective points (and their spatial spread) within each range-bin increases with increasing orientation angle of the reflective surface. As a result, the NLOS target is more severely masked by the echoes from the reflective surface, which degrades the estimation performance of the NLOS target parameters.
Notice that for high target SNR, the proposed approach achieves low $\mathrm{RMSE}$  of the target parameters estimation for all simulated reflective surface orientations, $\theta^w$.

\begin{figure}
     \centering
     \begin{subfigure}[b]{0.48\textwidth}
         \centering
         \includegraphics[width=\textwidth]{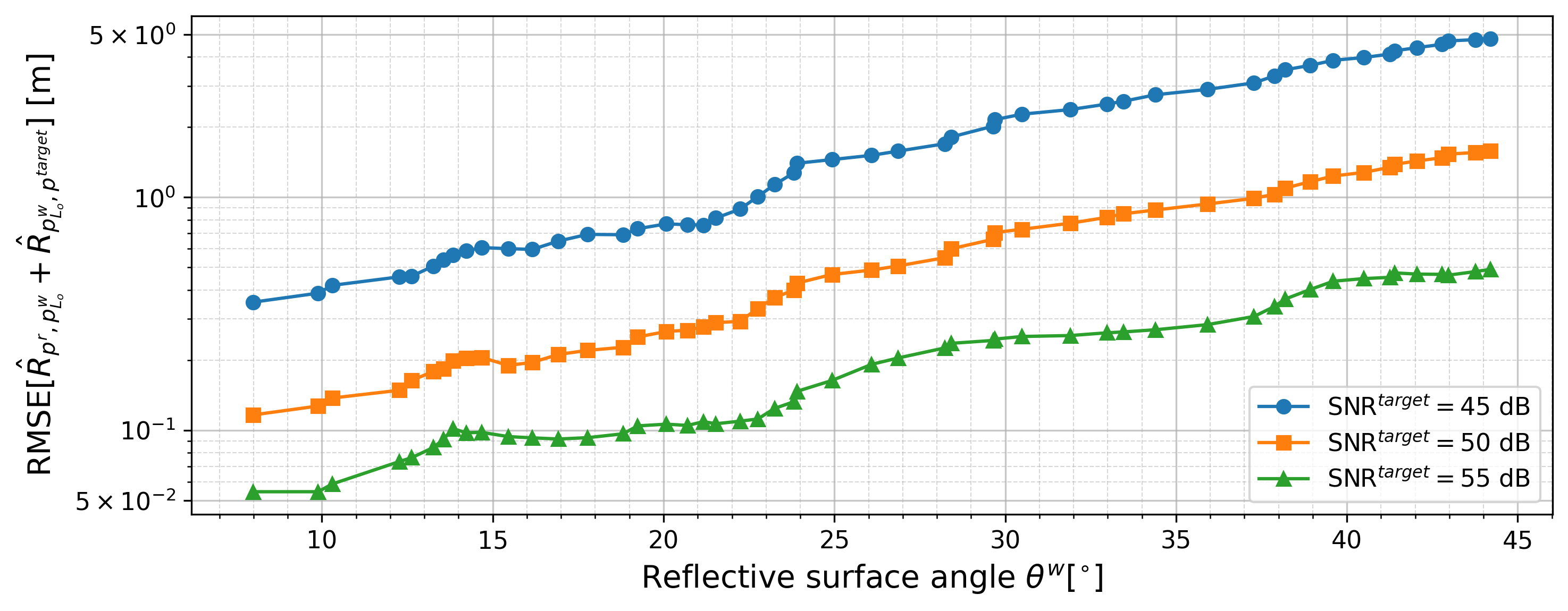}
         \caption{RMSE of $\bar{R}^*$ estimation as \\a function of $\theta^w$}    
         \label{fig:target_r_w_t_rmse_wall_angle_eval}
     \end{subfigure}
     \hfill 
     \begin{subfigure}[b]{0.48\textwidth}
         \centering
         \includegraphics[width=\textwidth]{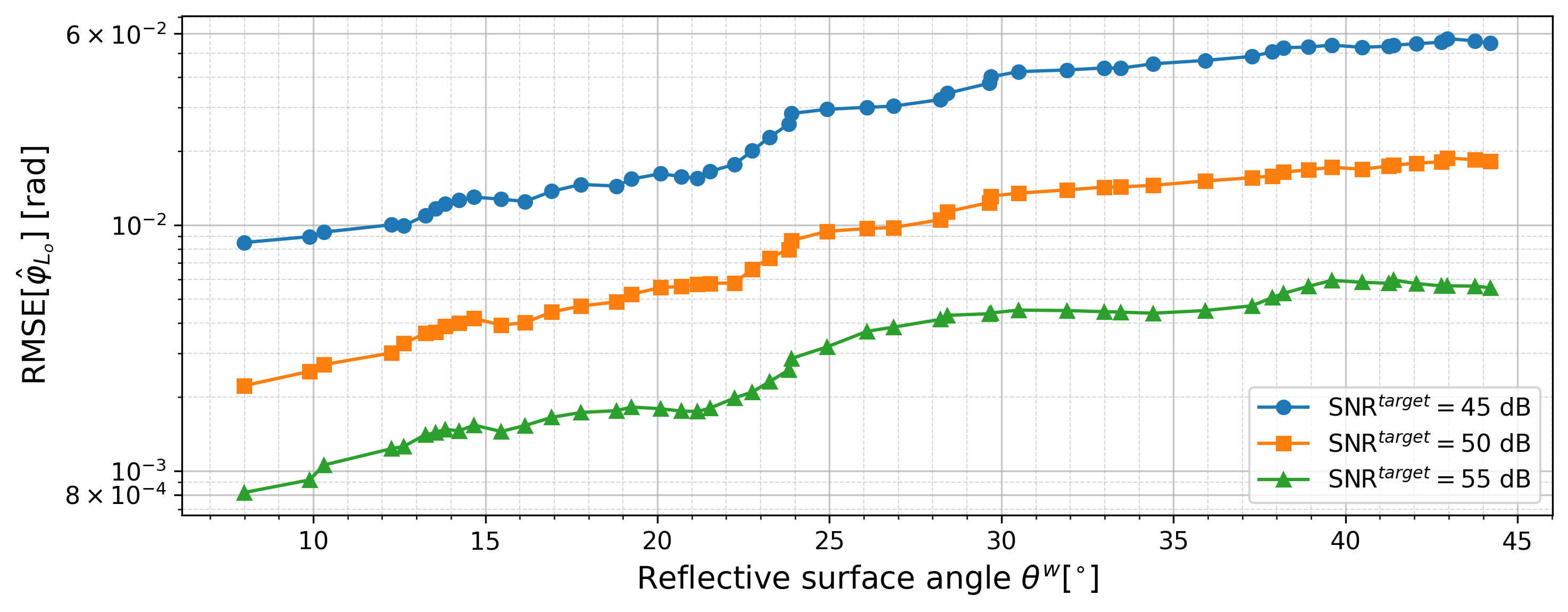}
         \caption{RMSE of $\hat{\varphi}_{K_o}$ estimation as a function of $\theta^w$.}
         \label{fig:target_theta_w_rmse_wall_angle_eval}
     \end{subfigure}
    \caption{\gls{RMSE} of target parameters estimation as a function of the reflective surface orientation angle, $\theta^w$, in scenarios with the target parameters, $\text{SNR}^{\text{\textit{target}}}=[45, 50, 55]\,\mathrm{dB}, \varphi_{K_o} = 9.46^\circ, R_{p^r, p^w_{K_o}} = 18.26m, {R}_{p^w_{K_o}, p^{\text{\textit{target}}}} = 11.89m$ and reflective surface parameters, $\text{SNR}^w=30\,\mathrm{dB}$, ${x}^w = 2m, {y}^w = 18m, {D}^w = 8m$. \textcolor{black}{Performance was evaluated using  $2.2 \times 10^4$ \gls{MC} simulations.}}    \label{fig:target_rmse_wall_angle_eval}
\end{figure}

Fig.~\ref{fig:target_radios_rmse_wall_angle_eval} shows the $\mathrm{RMSE}_d$ of the target localization in Cartesian coordinates, $(\hat{x}^{\text{\textit{target}}}, \hat{y}^{\text{\textit{target}}})$, as a function of the reflective surface orientation, $\theta^w$. 
The non-monotonic relation between the target parameters estimation $\mathrm{RMSE}_d$ and the reflective surface orientation, can be explained by the combination of the results in Figs. \ref{fig:wall_param_angle_eval} and \ref{fig:target_rmse_wall_angle_eval}, where the performance of reflective surface parameters estimation improves, and of the target parameters degrades with increasing reflective surface orientation, $\theta^w$.
Notice that for high target SNR, the proposed approach achieves a low $\mathrm{RMSE}_d$  of the target localization for all simulated reflective surface orientations, $\theta^w$.

\begin{figure}[htp]
    \centering 
    \includegraphics[width=0.48\textwidth]{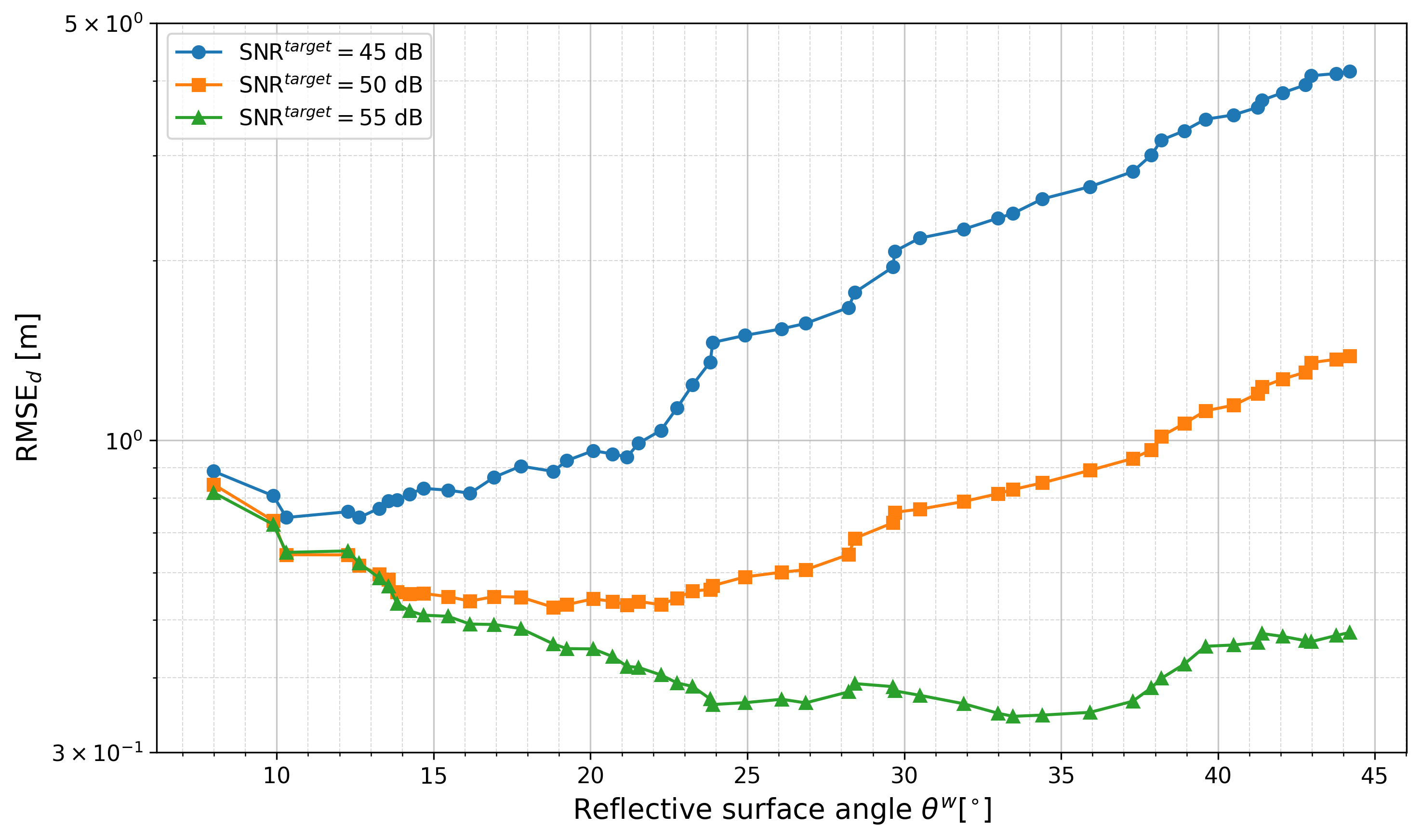}
    \caption{$\mathrm{RMSE}_d$ of target parameters estimation as a function of the reflective surface orientation angle, $\theta^w$, in scenarios with the target parameters, $\text{SNR}^{\text{\textit{target}}}=[45, 50, 55]\,\mathrm{dB}, \varphi_{K_o} = 9.46^\circ, R_{p^r, p^w_{K_o}} = 18.26m, {R}_{p^w_{K_o}, p^{\text{\textit{target}}}} = 11.89m$ and reflective surface parameters, $\text{SNR}^w=30\,\mathrm{dB}$, ${x}^w = 2m, {y}^w = 18m, {D}^w = 8m$. \textcolor{black}{Performance was evaluated using  $2.2 \times 10^4$ \gls{MC} simulations.}}
    \label{fig:target_radios_rmse_wall_angle_eval}
\end{figure}

\subsubsection{\texorpdfstring{Influence of the Reflective Surface Length, $D^w$}{Influence of the Reflective Surface Length, D w}}
 \label{sec:eval_wall_len}
Fig.~\ref{fig:target_wall_len_eval} shows that the performance of the $\bar{R}^*$ and $\hat{\varphi}_{K_o}$, estimation degrade with increasing reflective surface length, $D^w$ beyond $4.5$ [m].
Similarly to the results in subsection C 2), the \gls{RMSE} degradation with increasing reflective surface length, $D^w$, in Fig.~\ref{fig:target_wall_len_eval} can be explained by the higher sidelobes of the radar echoes from the reflective surface that mask the target, especially at low $\text{SNR}^{\text{\textit{target}}}$.
Notice that the influence of the reflective surface on the target estimation performance is more significant when the estimation of the reflective surface parameters is inaccurate. 
In such scenarios, masking the reflective surface can erroneously mask the target and degrade its localization performance. Fig.~\ref{fig:target_wall_len_eval} shows that increasing the reflective surface length initially improves the target parameters estimation performance due to improved $\text{SNR}^w$. However, a further increase in the reflective surface length results in the target masking and the degradation of the target localization performance.    

\begin{figure}
     \centering
     \begin{subfigure}[b]{0.48\textwidth}
         \centering
         \includegraphics[width=\textwidth]{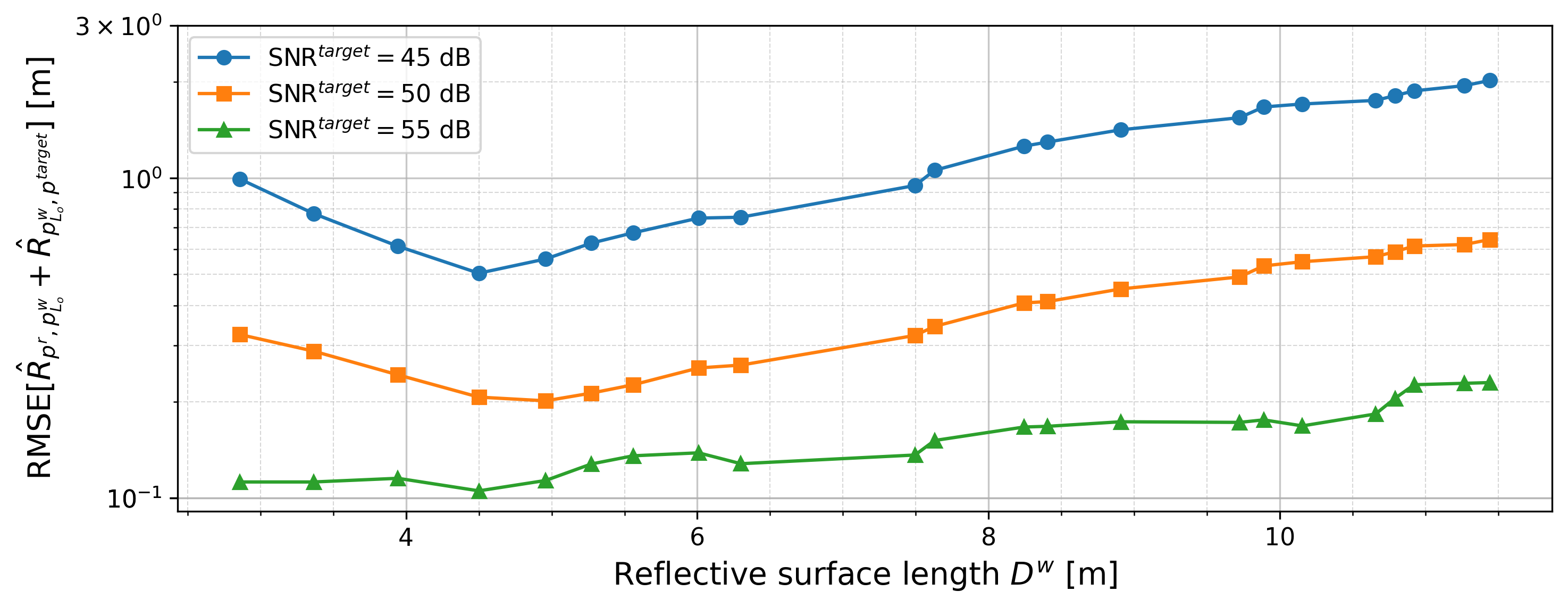}
         \caption{RMSE of $\bar{R}^*$ estimation as a function of the reflective surface length,  $D^w$.}    
         \label{fig:target_r_w_t_rmse_wall_len_eval}
     \end{subfigure}
     \hfill 
     \begin{subfigure}[b]{0.48\textwidth}
         \centering
         \includegraphics[width=\textwidth]{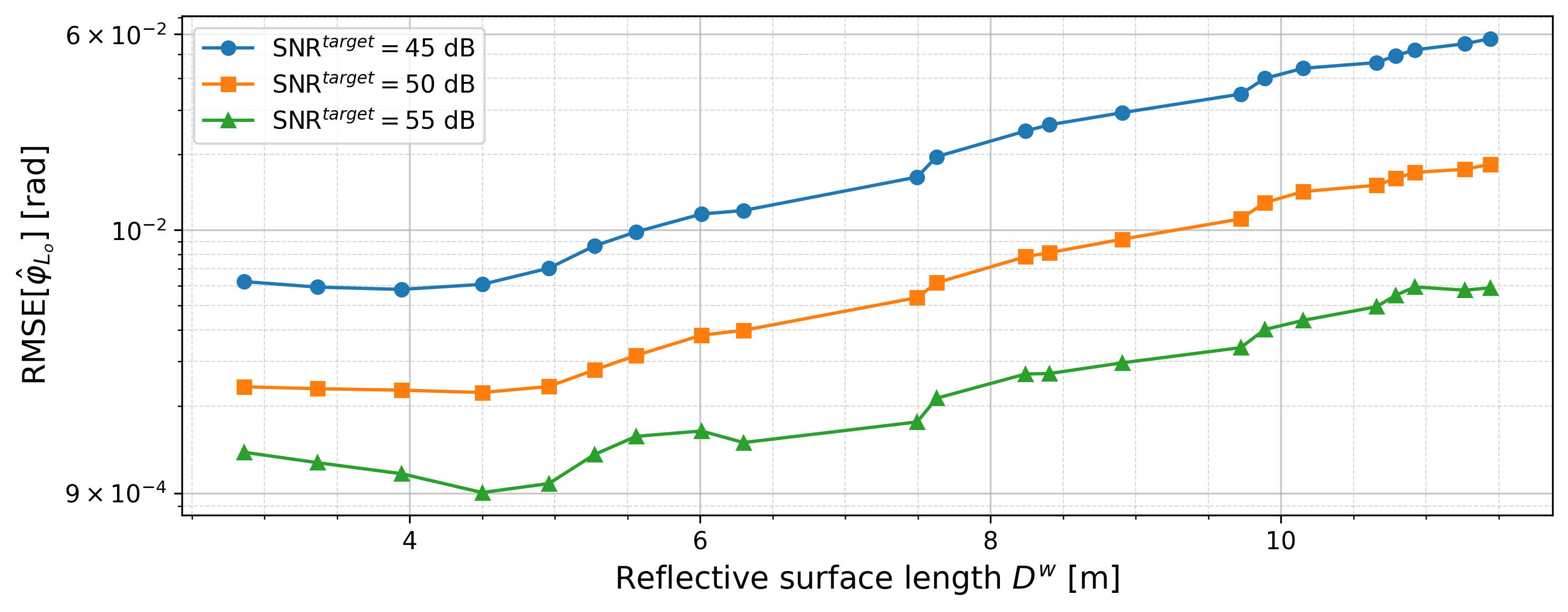}
         \caption{RMSE of $\hat{\varphi}_{K_o}$ estimation as a function of the reflective surface length, $D^w$.}
         \label{fig:target_theta_w_rmse_wall_len_eval}
     \end{subfigure}
    \caption{\gls{RMSE} of target parameters estimation as a function of the reflective surface length $D^w$ in scenarios with the target parameters, $\text{SNR}^{\text{\textit{target}}}=[45, 50, 55]\,\mathrm{dB}, \varphi_{K_o} = 9.46^\circ, R_{p^r, p^w_{K_o}} = 18.26m, {R}_{p^w_{K_o}, p^{\text{\textit{target}}}} = 11.89m$ and reflective surface parameters, $\text{SNR}^w=30\,\mathrm{dB}$, ${x}^w = 2m, {y}^w = 18m, \theta^w = 25^\circ$. \textcolor{black}{Performance was evaluated using  $4.6 \times 10^4$ \gls{MC} simulations.}}
     \label{fig:target_wall_len_eval}
\end{figure}

Fig.~\ref{fig:target_radios_rmse_wall_len_eval} shows the influence of the reflective surface length, $D^w$, on the $\mathrm{RMSE}_d$ of the target localization. 
Notice that the $\mathrm{RMSE}_d$ first improves with increasing length of the reflective surface due to improved performance of the reflective surface parameters estimation. However, a further increase in the reflective surface length degrades the $\mathrm{RMSE}_d$, due to masking weak targets by the strong echoes from the reflective surface.
All results demonstrate the robustness of the proposed approach to the variability of the considered scenario parameters.   
\begin{figure}[htp]
    \centering 
    \includegraphics[width=0.48\textwidth]{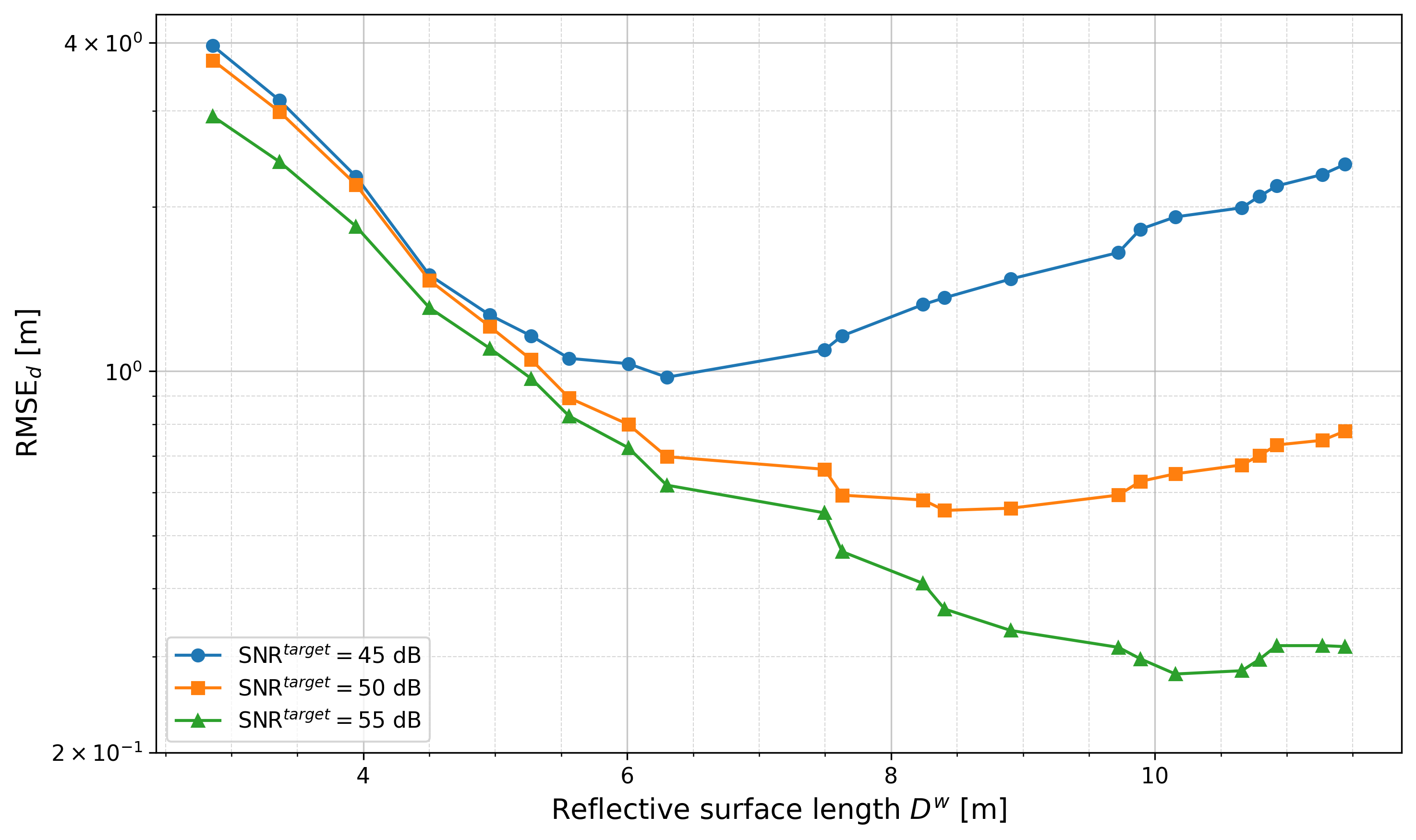}
    \caption{$\mathrm{RMSE}_d$ of target parameters estimation as a function of the reflective surface length $D^w$ in scenarios with target parameters, $\text{SNR}^{\text{\textit{target}}}=[45, 50, 55]\,\mathrm{dB}, \varphi_{K_o} = 9.46^\circ, R_{p^r, p^w_{K_o}} = 18.26m, {R}_{p^w_{K_o}, p^{\text{\textit{target}}}} = 11.89m$ and reflective surface parameters, $\text{SNR}^w=30\,\mathrm{dB}$, ${x}^w = 2m, {y}^w = 18m, \theta^w = 25^\circ$. \textcolor{black}{Performance was evaluated using  $4.6 \times 10^4$ \gls{MC} simulations.}}
    \label{fig:target_radios_rmse_wall_len_eval}
\end{figure}

\subsubsection{Reflective Surface's Model Mismatch}
This subsection evaluates the robustness of the proposed NLOS target localization approach to the mismatch in the considered model of the straight reflective surface.
The non-straight reflective surface with random irregularities is simulated using the following model: 
\begin{eqnarray}
    y^w_{k} = \Tilde{x}^w_{k} \tan(\theta^w) + b^w\;,\;\forall {k}=1,\ldots,{K}\;,
\end{eqnarray}
where $\Tilde{x}^w_{k} = x^w_k + \eta_x$, $\eta_x \sim \mathcal{N}(0,\, \sigma^2_x)$. 
Fig.~\ref{fig:missmatch_eval} shows the RMSE of the reflective surface parameters estimation as a function of the variance, $\sigma^2_x$, of the random deviations from the considered straight reflective surface model. 
Notice the only slight degradation in the RMSE of the reflective surface parameters estimations with growing deviations from the considered model of the straight reflective surface.  

\begin{figure}
     \centering
     \begin{subfigure}[b]{0.48\textwidth}
         \centering
         \includegraphics[width=\textwidth]{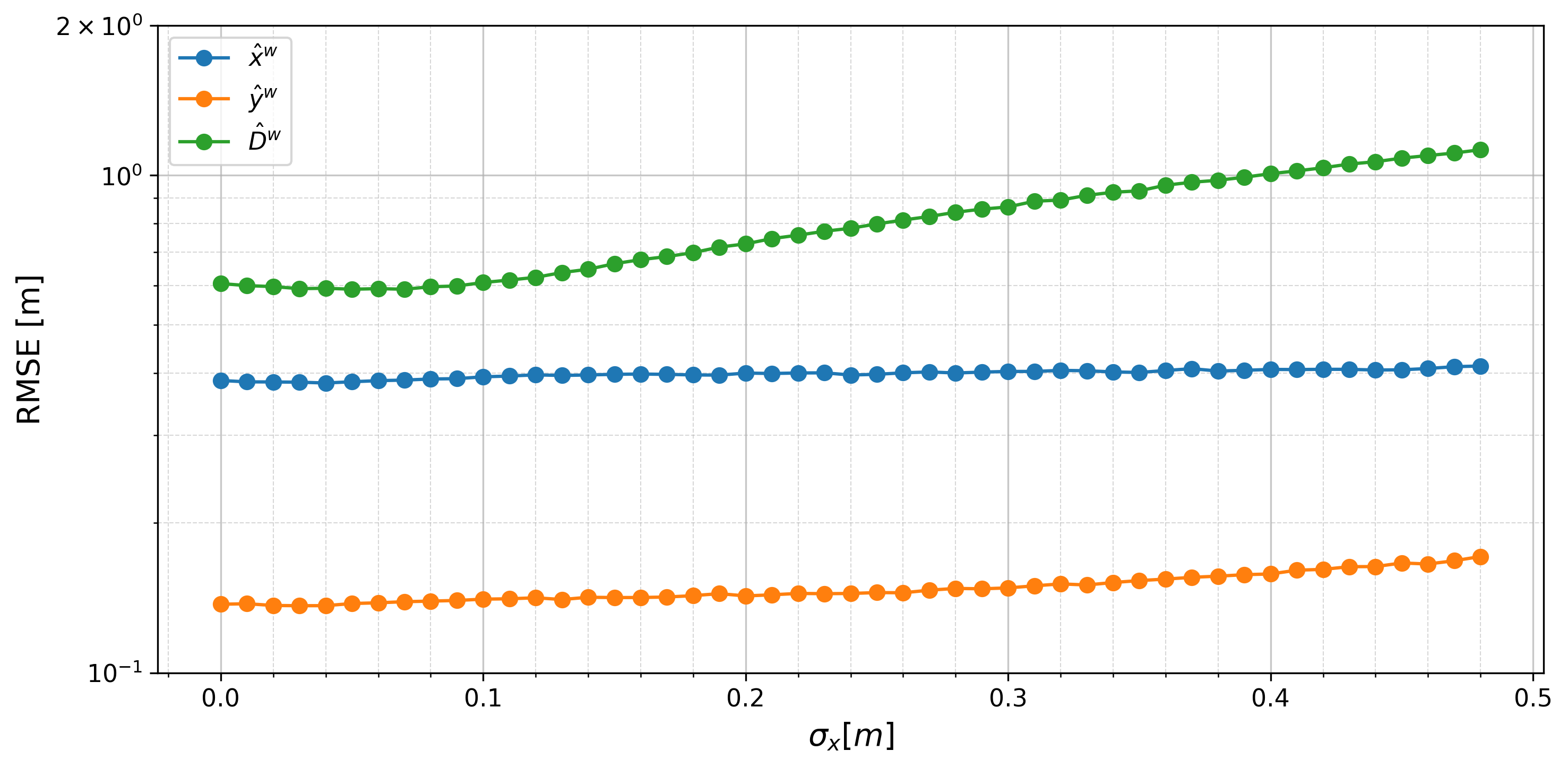}
         \caption{RMSE of $[\hat{x}^w, \hat{y}^w, \hat{D}^w]$ estimation as a function of $\sigma_x$ [m].}
         \label{fig:wall_x,y,length,_wall_missmatch_eval}
     \end{subfigure}
     \hfill 
     \begin{subfigure}[b]{0.48\textwidth}
         \centering
         \includegraphics[width=\textwidth]{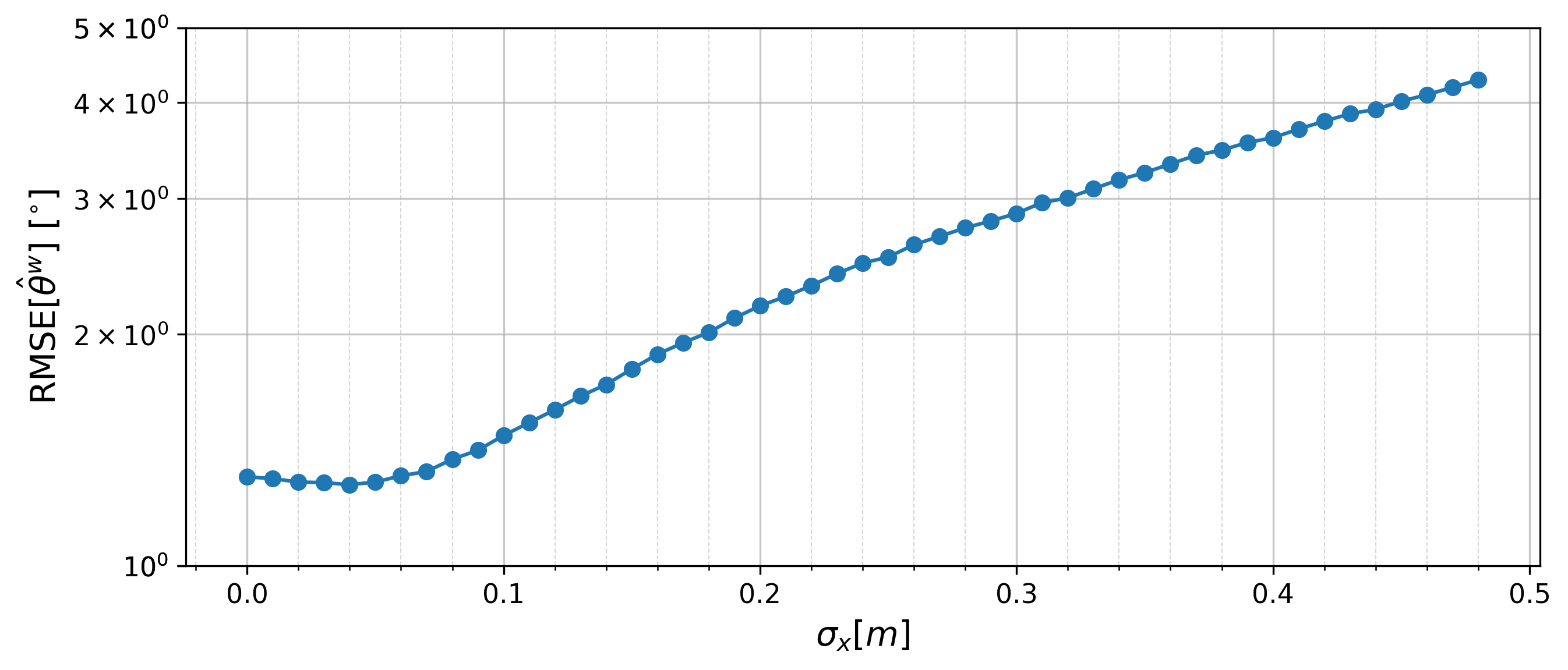}
         \caption{RMSE of $\hat{\theta}^w$ estimation as a function of $\sigma_x$ [m].}
         \label{fig:wall_angle_wall_missmatch_eval}
     \end{subfigure}
\caption{\gls{RMSE} of reflective surface parameters estimation as a function of $\sigma_x$ in scenarios with the following reflective surface parameters: $\text{SNR}^w=30\,\mathrm{dB}$, ${x}^w = 2m, {y}^w = 18m, \theta^w = 25^\circ, D^w=8m$. \textcolor{black}{Performance was evaluated using  $2.6 \times 10^4$ \gls{MC} simulations.}}
\label{fig:missmatch_eval}
\end{figure}
Fig.~\ref{fig:target_radios_rmse_wall_missmatch_eval} shows the influence of the model mismatch on the  $\mathrm{RMSE}_d$ of the target parameters estimation. Notice that the $\mathrm{RMSE}_d$ increases linearly with $\sigma_x$, remaining below $10\%$ of the target range, $R_{p^r, p^w_{K_o}} + {R}_{p^w_{K_o}, p^{\text{\textit{target}}}}$, for all simulated $\sigma_x$ values.

\begin{figure}[htp]
    \centering 
    \includegraphics[width=0.48\textwidth]{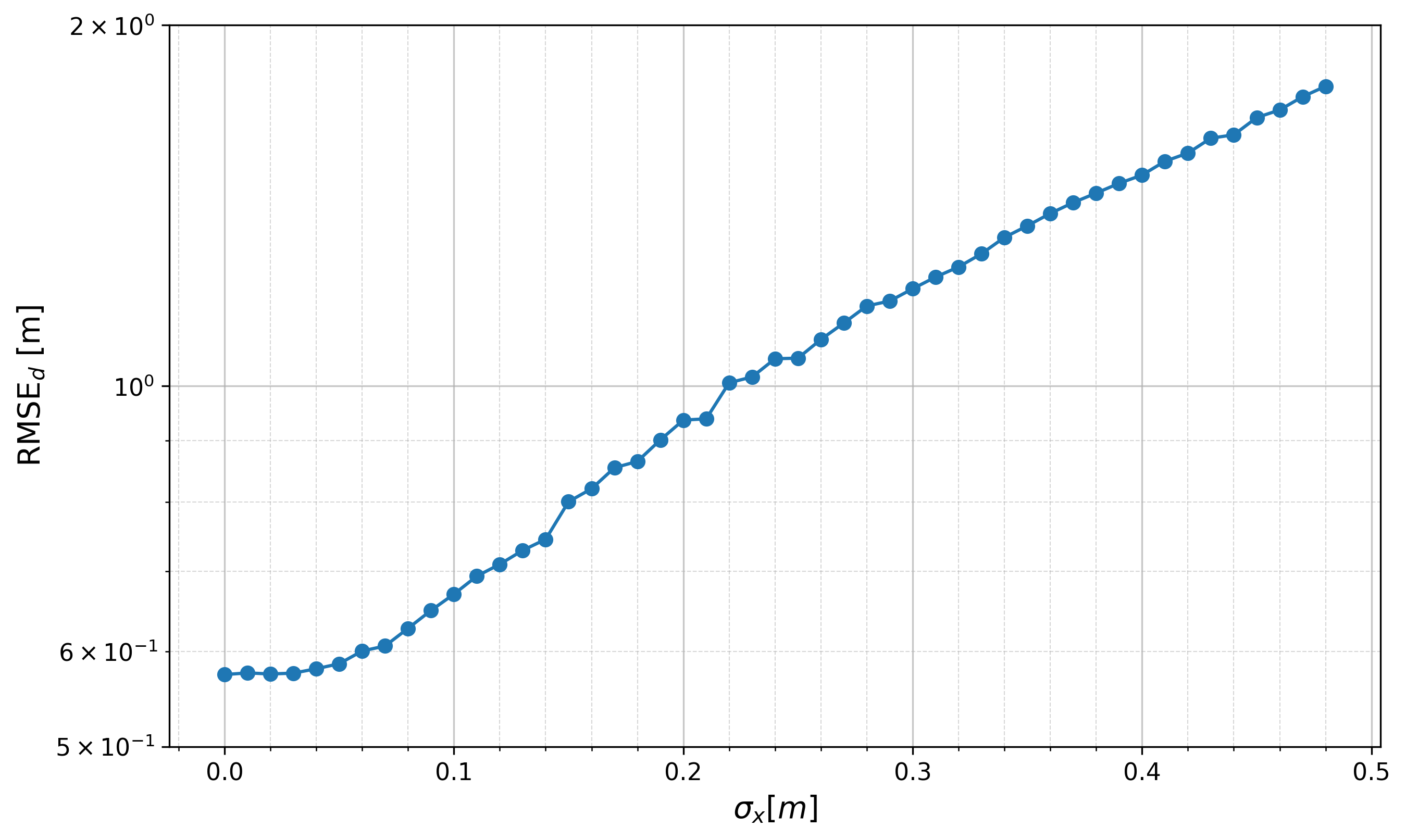}
    \caption{$\mathrm{RMSE}_d$,  of target parameters estimation as a function of $\sigma_x$ [m] in scenarios with the following reflective surface parameters: $\text{SNR}^w=30\,\mathrm{dB}$, ${x}^w = 2m, {y}^w = 18m, \theta^w = 25^\circ, D^w=8m$, and target parameters are: $\varphi_{K_o}=6.3^\circ, R_{p^r, p^w_{K_o}} = 18.1m, {R}_{p^w_{K_o}, p^{\text{\textit{target}}}} = 11.9m$. \textcolor{black}{Performance was evaluated using  $2.6 \times 10^4$ \gls{MC} simulations.}}
    \label{fig:target_radios_rmse_wall_missmatch_eval}
\end{figure}

\subsection{Computational Complexity}
The computational complexity of the proposed approach is evaluated in this section considering {\it ConViT}, {\it EfficientNet $b_0$}, and {\it EfficientNet $b_1$} CNN architectures in $\mathbf{C}(\cdot)$ layer of the proposed approach. Table~\ref{table:complexity} shows the average processing time of the proposed approach with the considered architectures \textcolor{black}{, averaged over $10^5$ \gls{MC} simulations.}.

The proposed approach with the {\it EfficientNet $b_1$} has a frame rate of $\sim 100$Hz, which, in automotive applications, can be considered real-time. It can further be optimized and implemented using dedicated hardware. Notice that the alternative architectures, {\it ConViT} and {\it EfficientNet $b_0$}, are faster, achieving a frame rate of $\sim 135$Hz, but provide lower performance. 
\begin{table}[!ht]
    \centering
    \caption{Computational complexity of the proposed approach using various CNN architectures in $\mathbf{C}(\cdot)$ layer, measured using AMD Ryzen Threadripper PRO 5965WX, with an Nvidia RTX A5000 Ada GPU.}
       \label{table:complexity}
    \begin{tabular}{m{1.8cm} m{1.8cm} m{1.5cm} m{1.5cm}}
    \hline\hline
        $\mathbf{C}(\cdot)$ layer & Runtime relative& $\mathbf{H}(\cdot)$ Total & Average \\
         model type & to {\it EfficientNet $b_1$} &\#Parameters & runtime\\\hline\hline
        \textbf{\textit{EfficientNet} $\boldsymbol{b_1}$} & \textbf{1x} & \textbf{7.8M} & \textbf{9.54 [ms]} \\ \hline
        {\it EfficientNet $b_0$} & 0.78x & 5.3M & 7.43 [ms] \\ \hline
        {\it ConViT} & 0.77x & 6.7M & 7.38 [ms] \\ \hline
    \end{tabular}
\end{table}

\section{Conclusions}
This work addresses the challenging problem of automotive radar NLOS target localization. It proposes a novel hybrid approach for \gls{LOS}/\gls{NLOS} radar target localization, combining deep learning processing with a physical model of electromagnetic wave propagation. Unlike conventional approaches that rely on prior knowledge of the environment or auxiliary sensors, the proposed approach accurately estimates reflective surface parameters and identifies the LOS/NLOS propagation conditions using only radar data.  
The key novelty of this approach lies in its ability to achieve accurate NLOS target localization without requiring predefined environmental maps or additional sensors. The proposed hybrid approach significantly outperformed conventional LS and RANSAC. The hybrid framework also demonstrated superior LOS/NLOS identification performance, with robust performance in all considered surface orientations and lengths. The robustness of the proposed approach to the nonlinearity of the considered reflective surface was demonstrated.
The proposed approach extends the automotive radar operational capabilities in dense urban environments and, as a result, can enhance the safety and reliability of autonomous driving systems.

\textcolor{black}{This work establishes the foundational framework for future research. Specifically, for clarity and tractability, a point-target model was considered in this work. Extending the proposed NLOS targets localization framework to handle practical extended targets, by modeling them as a superposition of multiple point targets, is a natural and straightforward extension, which is a subject of our future work. Further, our future work will consider various practical, complex urban scenarios, such as intersections, turns, and others, which are characterized by multiple complex objects and reflective surfaces. }

\section*{Acknowledgement}
The authors thank Dr. Inna Stainvas from GE Healthcare Science $\&$ Technology for her expertise, insightful discussions, and productive ideas.  
\bibliographystyle{IEEEtran}
\bibliography{bibliography}
\begin{IEEEbiography}[{\includegraphics[width=1in,height=1.25in,clip,keepaspectratio]{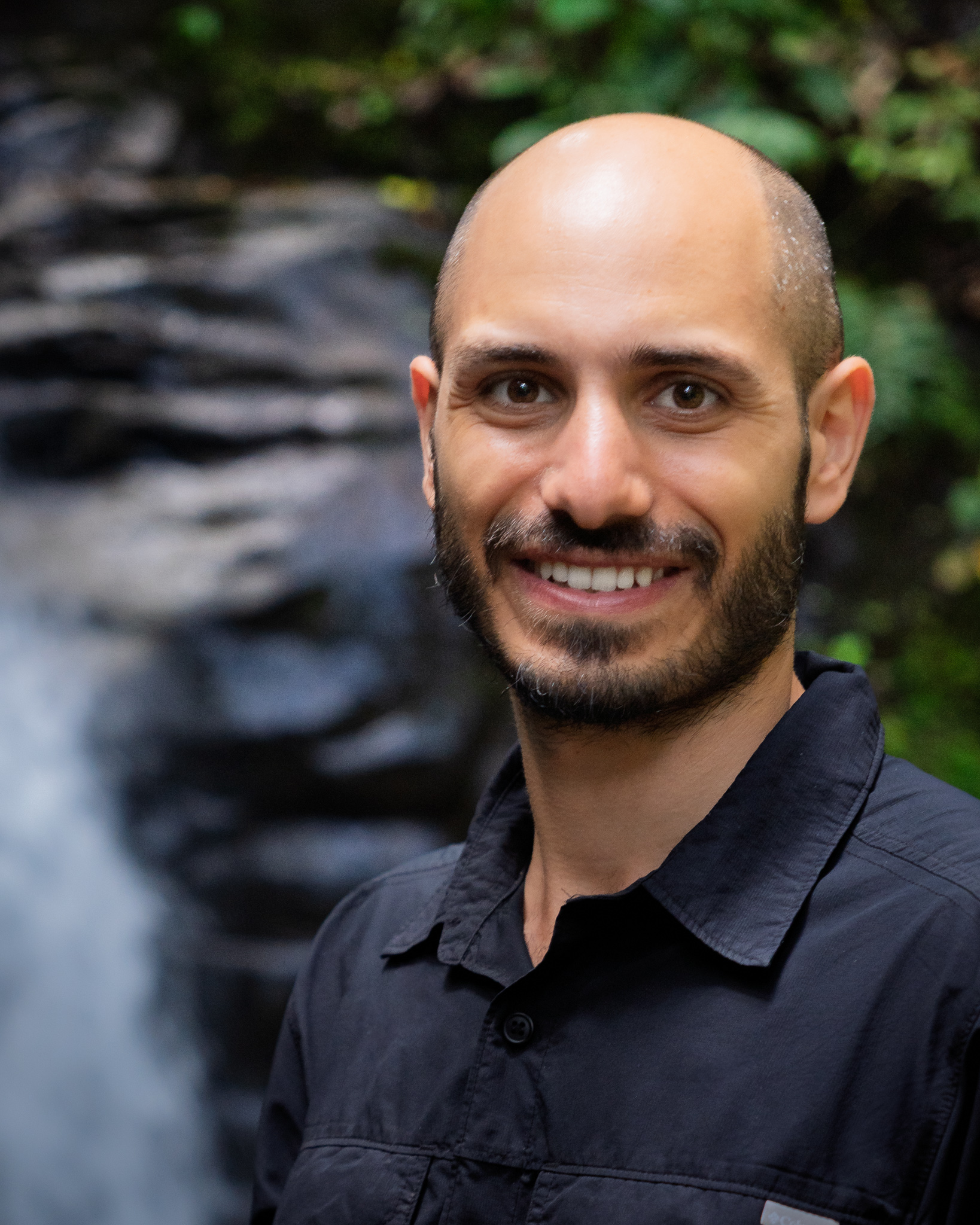}}]%
{Aviran Gal} received the B.Sc. and M.Sc. degrees in electrical and computer engineering from Ben-Gurion University of the Negev, Beer Sheva, Israel, in 2021 and 2024, respectively. From 2019 to 2022, he was employed by DSP Group, where he was involved in the development of voice processing and wireless chipset solutions. He is currently with the Department of Electrical and Computer Engineering, Ben-Gurion University of the Negev, Beer Sheva, Israel, as an Algorithm Developer focusing on advanced signal processing techniques. His research interests include signal processing for automotive radar, with an emphasis on combining deep learning and statistical signal processing approaches.
\end{IEEEbiography}
\begin{IEEEbiography}[{\includegraphics[width=1in,height=1.25in,clip,keepaspectratio]{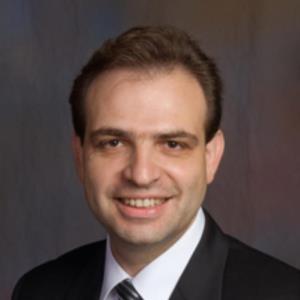}}]%
{Igal Bilik}(S'03-M'06-SM'21) received B.Sc., M.Sc., and Ph.D. degrees in ECE from Ben-Gurion University of the Negev in 1997, 2003, and 2006, respectively. During 2006–2008, he was a postdoctoral research associate in the Department of ECE at Duke University. During 2008-2011, he was an Assistant Professor in the Department of ECE at the U. of Massachusetts, Dartmouth. During 2011-2020, he was a Staff Researcher and Smart Sensing and Vision Group Leader at GM Advanced Technical Center. Since 2020, Dr. Bilik has been an Assistant Professor in the School of Electrical and Computer Engineering at Ben-Gurion University of the Negev. He has been a member of the IEEE AESS Radar Systems Panel Committee and a chair of the Civilian Radar Committee. Dr. Bilik is an Acting Officer of the IEEE Vehicular Technology Chapter in Israel and vice-chair of the IEEE AESS Chapter in Israel. He is an Associate Editor-in-Chief for IEEE TAES and an AE of the IEEE Sensors and IEEE TRS. He is a Member of the Transactions on Radar Systems Editorial Committee. Dr. Bilik received the Best Student Paper Awards at IEEE RADAR 2005 and IEEE RADAR 2006 Conferences, the Student Paper Award at the 2006 IEEE 24th Convention of Electrical and Electronics Engineers in Israel, the GM Product Excellence Recognition in 2017, and the IEEE TAES Industrial Innovation Award in 2024. Currently, he is an IEEE AESS Distinguished Lecturer. 
\end{IEEEbiography}

\end{document}